\documentstyle[twocolumn,prc,aps,epsfig]{revtex}
\input epsf
\newcommand{\be}{\begin{eqnarray}}
\newcommand{\ee}{\end{eqnarray}}

 \newcommand{\gsim}{\mathrel{\hbox{\rlap{\lower.55ex \hbox {$\sim$}}
                   \kern-.3em \raise.4ex \hbox{$>$}}}}
\newcommand{\lsim}{\mathrel{\hbox{\rlap{\lower.55ex \hbox {$\sim$}}
                   \kern-.3em \raise.4ex \hbox{$<$}}}}

\def\roughly#1{\mathrel{\raise.3ex\hbox{$#1$\kern-.75em%
\lower1ex\hbox{$\sim$}}}}
\def\lsim{\roughly<}
\def\gsim{\roughly>}

\def\la{{\Big<}}
\def\ra{{\Big>}}

\begin{document}

\twocolumn[\hsize\textwidth\columnwidth\hsize\csname @twocolumnfalse\endcsname

\title {
{Instanton-induced Effects in QCD High-Energy Scattering  } 
}
\author {
 Edward Shuryak and Ismail Zahed
}
\address {
 Department of Physics and Astronomy\\ State University of New York, 
     Stony Brook, NY 11794-3800 
}
\date{\today}
\maketitle    
\begin{abstract}

We evaluate a number of new instanton-induced phenomena in QCD, starting
with static dipole-dipole potentials, and proceeding to 
quark-quark and dipole-dipole scattering at high energy. We use a non-perturbative formulation of the scattering amplitude in terms of
a correlator of two Wilson-lines (quarks) or Wilson-loops (dipoles) and 
analyze the Euclidean amplitudes with both perturbative gluons
and  instantons. The results are  analytically continued  to 
Minkowski geometry, by interpreting the angle between the Wilson lines 
as rapidity. We discuss the  relevance of our results for the phenomenology 
of near-forward hadronic processes at high energy, especially for
processes
with multiple color exchanges.
\end{abstract}
\vspace{0.1in}
]
\begin{narrowtext}   
\newpage

















\section{Introduction}\label{intro}
\indent
Significant progress reached in the realm of non-perturbative QCD 
has been mostly related with approaches based on the Euclidean
formulation of the theory: numerical simulations using lattice
gauge theory, instantons, monopoles, etc. By now, we know a great
deal about the important or even dominant role of instanton-induced effects for
correlation functions in a variety of hadronic channels,
hadronic wave functions and form-factors, for a review
see~\cite{SS_98}. 
Unfortunately so far
 many of those results have not been translated to Minkowski 
space, a crucial step for understanding hadronic high-energy processes.
It is however clear that there must be a very general and direct
relationship between the hadronic substructure and the details 
of high energy  reactions. Indeed,  the non-perturbative modification of
QCD vacuum fields induced by the valence quarks studied in
Euclidean space-time should 
look like parton correlations in the transverse plane in a boosted
frame. 
 Many known features of partonic distributions,
including spin and flavor of the sea quarks, point to their
non-perturbative origin. Many more features (like 
$fluctuations$ of these cross sections and $correlations$
in the parton positions in the transverse plane which we briefly 
discuss at the  end of the paper) are still to be studied in details.

The first  systematic  step towards a semi-classical but
non-perturbative formulation of high-energy scattering
in QCD was suggested by Nachtmann~\cite{NACHTMANN},
who has related the scattering amplitude  to 
expectations of pairs of Wilson lines.
Semi-classical expressions with a similar pair of Wilson lines
for DIS structure functions were also proposed by
Muller \cite{Muller}: in contrast to their traditional
interpretation as partonic densities, they were treated as
cross sections for targets penetrated by small dipole-like
probes at high energy. 
One systematic way to use these semi-classical expressions
is to go back to the perturbative domain and try to
improve on the diagrammatic approaches (like the celebrated BFKL~\cite{BFKL}
re-summation): see e.g. calculations of the anomalous
dimension of the cross-singularity between two Wilson
lines~\cite{KORCHEMSKY} or  the
analysis of the path-exponents in \cite{Balitsky}.

The approach we will follow in this paper is different: 
the  Wilson lines in question are evaluated semi-classically
using instantons. In order to be able to do so, one should start in
Euclidean space-time, where those solutions are  the saddle points 
of the functional integrals.  The results are then analytically continued
back to Minkowski space. Although it was not done before in this form, 
there are similar approaches in the perturbative context (e.g.
\cite{MEGGIOLARO} and references therein). 
Another methodically close approach to our analysis
is~\cite{RSZ,JANIK} where recent progress on the non-perturbative dynamics 
in N=4 SYM theory was used. In particular, the AdS/CFT correspondence 
have been used to evaluate the partonic cross section geometrically, 
using a deformed string in the curved anti-de-Sitter space.

   The instanton-induced processes to be considered in this work are either
$elastic$ scattering of partons, or {\it quasi-elastic} ones, with 
color transfer between them. They are very different from (and
should not be confused with) multi-quanta production processes originally
discussed
in electroweak theory \cite{Ringwald:1990ee} in connection with baryon number violation
and later in QCD in connection to DIS \cite{Ringwald_HERA}. Such
phenomena,
associated with small-size instantons, are easier to evaluate and also they
should lead to much more spectacular events. However, those lead to
much smaller cross sections
in comparison to the processes 
to be discussed below.

In this paper, we will not aim at  a development of a realistic model
for high-energy hadronic reactions based on instanton physics.
Instead, we will  answer few questions of principle, 
such as : Is it possible to assess non-perturbatively 
scattering amplitudes using the Euclidean formulation of
the theory? How is the analytical continuation enforced
on the non-perturbative amplitudes? What is the magnitude of
the instanton induced effects in comparison to the perturbative
effects in the scattering of near-forward high energy partons?

In section 2 we review the perturbative effects on the
dipole-dipole potential, including the derivation of
a renormalization group solution that can be tested 
using QCD lattice simulations. In section 3 we
extend the perturbative analysis in Euclidean space to
the case of scattering between two quarks and two
dipoles. Particular issues regarding the analytical
continuation of the perturbative results to Minkowski 
space are discussed.  In section 4, we discuss the
effects of instantons on the static potentials
for quarks and dipoles. At large distances the results
resemble perturbation theory apart from the large
classical enhancement of $(8\pi^2/g^2)^2\approx 10^2$,
which is partially compensated by the diluteness factor
$n_0\rho_0^4\approx (1/3)^4$~\cite{Shu_82} of the instantons in the  vacuum. 
In section 5, we  calculate the scattering amplitudes for
quarks and dipoles in the one-instanton approximation.
The color preserving part of the amplitude is real and vanishes
at high energy. The color exchange part is real 
but finite at high energy, thereby contributing to the
near-forward inelastic scattering or re-scattering of partons. 
In section 6, we extend our discussion to two-instantons.
We found  that for two quarks the cross section
is of the order of $\sigma_{qq}\sim  (n_0\rho_0^4)^2 \rho_0^2$, while for
two dipoles it is  further suppressed $  \sigma_{dd}\sim \sigma_{qq}
(d^2_1 d^2_2/\rho_0^4)$. These results are supported by our calculations.
In section 7, we discuss the possible role of instantons in cross-section
fluctuations. Our conclusions and recommendations are in section 8.

\section{Perturbative  Analysis of Potentials}\label{QQPT}

\subsection{Dipole- dipole potential}
We start with the simplest analysis in Euclidean space, in which the
perturbative expansion of two  Euclidean Wilson lines leads to the
well-known result for the potential between static charges.
Indeed, by expanding two Wilson lines to first order in the
gauge-coupling $g$, using the  Euclidean propagator 
$\left<A\,(x) A\,(y)\right>\sim   1/(x-y)^2$ with $x,y$  
located on two parallel but straight lines, and finally
integrating over the relative time $x_0-y_0$, we readily obtain
the Coulomb potential $V(R)\sim \alpha_s(R)/R$.

Now, consider the case of the interaction between two 
{\em color neutral} objects, such as {\it two static color dipoles}.
The simplest perturbative process
 in this case includes
 double photon/gluon exchange. 
 The problem was solved in QED 
by Casimir and Polder \cite{CP}, who have shown
that the  potential at large
distances R 
is 
\be V(R)=- {\alpha_1  \alpha_2 \over R^7} \ee
where the {\it polarizabilities} $\alpha_{1,2}$  are of the
 order of
 $\alpha_{1,2}\sim \tau_0 d^2$,
$d$ is the dipole size and  $\tau_0$ is some characteristic time
(see below). 
This result differs from the  Van-der-Waals potential
 $1/R^6$ (valid at smaller R)
because of the time delay effects. These observations were generalized
to perturbative QCD in \cite{PESKIN,Kharzeev_pot}.

 The Euclidean approach leads to the  7-th power of R in a 
  simple way, provided that  
the following conditions are satisfied:
 (i) $d_{1,2} \ll  R$ which justifies the dipole approximation and
 identifies the relevant
field operators
 $(\vec d\cdot \vec E)^2$; (ii)   $both$ exchanged photons (or gluons) are emitted and
absorbed at close  $x_0$ and $y_0$ times. As a result, the 
perturbative  field correlator
$<E^2(x)E^2(y)>\sim 1/(x-y)^8=1/(R^2+\tau^2)^4$, once  integrated over 
 the relative time $\tau=x_0-y_0$, leads the result $1/R^7$.

The  condition (ii) can be  understood
for complex systems like atoms or hadrons in the following way:
the first dipole emission excites the system from (usually an S-wave)
ground state to (usually a P-wave) excited state, 
while the second  dipole emission returns the system back.
The energies of the intermediate state sets
the characteristic life-time
$\tau_0 \approx 1/(E_P-E_S) $.

 However for  {\it static} dipoles the situation is  different
 in QED and QCD.
In QED the emission times
of two exchanged quanta are independent, but in QCD they are not.
Even  a {\it static} dipole
can change its color degrees of freedom. Because
 different total color
states of the dipole have different energies, thanks to the
Coulomb interaction, we again have excited intermediate states.
Therefore the characteristic time is
determined by the difference in Coulomb energy between
the singlet and octet states 
\be 1/\tau_0= \Delta E= (3 \alpha_s/ 2)/ d \,\,.
\ee
Although the
dipoles may be $small$ $d\ll R$, this time
may still be long because in the perturbative domain 
the coupling constant is small $g^2(d)\ll 1$.
 As a result,  there are two  different
regimes,  when the
distance R is large (i) $R\gg \tau_0$ or small (ii)  $R\ll \tau_0$.
In the former case again the power is 7 and
the polarizability is\footnote{Amusingly,
the result is just the volume of a sphere of radius d,
from which the perturbative coupling constant 
$g$ dropped out.}  
$\alpha=4\pi d^3/3$. The latter case is the Van-der-Walls domain.

\subsection{RGE analysis of the dipole-dipole potential}

On general grounds, 
the potential between two interacting dipoles
can be shown to obey the following equation 
\begin{equation}
\alpha_s\, \frac{\partial{\cal V} (b)}{\partial \alpha_s} = -\frac 12 \int\, d^3x \, \left< {\rm Tr} \, F^2 (x)\,\right>_b
\label{RGE1}
\end{equation}
where $\alpha_s$ is the QCD running coupling and the averaging in (\ref{RGE1}) is carried
in the presence of the two static dipoles a distance $b$ apart.
Generically,
\begin{equation}
{\cal V} (b) \equiv {\cal V} (b, a, \mu , \alpha_s) \approx \mu \, (\mu \, a)^\kappa\, F (\mu\, b, \alpha_s) 
\label{RGE2}
\end{equation}
where $\mu$ is the renormalization scale.
Hence,
\begin{equation}
\frac{\partial{\cal V}}{\partial \alpha_s} =-\frac 1{\beta}
\left((\kappa + 1) \, {\cal V} + b\,\frac{\partial{\cal V}}{\partial b}\right)
\label{RGE3}
\end{equation}
with $\beta=d\alpha_s/d{\rm ln}\mu$ is the QCD beta function. Inserting (\ref{RGE3}) into
(\ref{RGE1}) yields
\begin{equation}
(\kappa + 1) \,{\cal V} + b\frac{\partial\,{\cal V}}{\partial b}=\frac {\beta}{2\alpha_s}\,\int\, d^3x \, 
\left<{ Tr} \, F^2 (x) \right>_b\,\,,
\label{RGE4}
\end{equation}
which is the RGE equation satisfied by the dipole-dipole potential. 
At large separations we may assume the dipole-dipole potential in quenched QCD to follow like a power 
law, i.e. 
\begin{equation}
{{\cal V} (b)}  \approx \mu\,(\mu a)^\kappa\,(\mu b)^\gamma \,,
\label{RGE5}
\end{equation}
turning (\ref{RGE4}) into an algebraic equation
\begin{equation}
(1+\gamma +\kappa )\, {\cal V} (b) = \frac {\beta}{2\alpha_s}\,\int \, d^3 x\, \left<{\rm Tr}F^2 (x)\right>_b\,.
\label{RGE6}
\end{equation}
Alternatively, the potential between two dipoles is a measure of the energy density 
in the presence of two-dipoles
\begin{equation}
{\cal V} (b) = \int\, d^3x\, \left<\Theta_{00} (x) \right>_b\,\,.
\label{RGE7}
\end{equation}
The combination of the RGE equation (\ref{RGE6}) and the definition (\ref{RGE7}) yields
a constraint between the exponents $\kappa$ and $\gamma$ in (\ref{RGE5}) asymptotically,
namely
\begin{equation}
\gamma =-1-\kappa +\frac {\beta}{\alpha_s}\,\frac {1-R}{1+R}
\label{RGE8}
\end{equation}
with 
\begin{equation}
R=\frac{\int\, d^3x \left< B^2 (x)\right>_b}{\int\, d^3x \left< E^2 (x)\right>_b}
\label{RGE9}
\end{equation}
a measure of the magnetic-to-electric ratio in the configuration composed
of two static dipoles a distance $b$ away from each other. For a self-dual
field $R=1$ and $\gamma=-1-\kappa$ if the asymptotic (\ref{RGE5}) is assumed.

\section{Perturbative Scattering In Euclidean Geometry}

\subsection{Quark-Quark scattering}

Generically, we will refer to  quark-quark scattering as
\be
Q_A(p_1) + Q_B(p_2) \rightarrow Q_C(k_1) +Q_D(k_2)
\label{3}
\ee
 We denote by
$AB$ and $CD$ respectively, the incoming and outgoing 
color and spin of the quarks (polarization for gluons). Using
the eikonal approximation and LSZ reduction,
the scattering amplitude ${\cal T}$ for quark-quark 
scattering reads~\cite{NACHTMANN,KORCHEMSKY,VERLINDE}
\begin{eqnarray}
&&{\cal T}_{AB,CD} (s,t) \approx 
-2is \int d^2b\,\, e^{iq_{\perp}\cdot b}\nonumber\\&&\times
\la \left({\bf W}_1 (b) -{\bf 1}\right)_{AC} 
\left( {\bf W}_2 (0) -{\bf 1}\right)_{BD}\ra
\label{4}
\end{eqnarray}
where as usual $s=(p_1+p_2)^2$, $t=(p_1-k_1)^2$, $s+t+u=4m^2$ and
\be
{\bf W}_{1,2}(b)= {\bf P}_c {\rm exp}\left(ig\int_{-\infty}^{+\infty}\,d\tau\,
A(b+v_{1,2}\tau)\cdot v_{1,2}\right)
\label{5}
\ee
The 2-dimensional integral in (\ref{4}) is over the impact parameter $b$
with $t=-q_{\perp}^2$, and  the averaging is over the gauge configurations
using the QCD action. The color bearing amplitude 
(\ref{4}) allows for scattering into a singlet or
an octet configuration, i.e. 
\be
{\cal T}=  {\cal T}_1\,\,{\bf 1}\otimes {\bf 1}  + {\cal T}_{N_c^2-1}\,\,({\tau^a}\otimes {\tau^a})
\label{8}
\ee
following the decomposition $N_c\otimes N_c =1 \oplus (N_c^2-1)$.
For gluon-gluon scattering the lines are doubled in color
space (adjoint representation) and further gauge-invariant 
contractions are possible. For quark-quark scattering the singlet
exchange in t-channel is 0$^+$ (pomeron) while for quark-antiquark
it is 0$^-$ (odderon) as the two differ by charge conjugation.

A quark with large momentum $p$
travels on a straight line with 4-velocity 
$\dot{x}=v=p/m$ and $v^2=1$. In the the eikonal approximation 
an ordinary quark transmutes to a scalar quark. The  argument applies to 
any charged particle in a background gluon field, with the following 
amendments: for anti-quarks the 4-velocity $v$ is 
reversed in the Wilson line and for gluons the Wilson lines are in the 
adjoint representation. Quark-quark scattering can be
also extended to quark-antiquark, gluon-gluon or scalar-scalar scattering.
For quark-antiquark scattering the elastic
amplitude dominates at large $\sqrt{s}$ since the annihilation
part is down by $\sqrt{-t/s}$.

It can be described in
Minkowski geometry 
in the CM frame with
$p_1/m= \,({\rm cosh} {\gamma}/2, \,{\rm sinh} {\gamma}/2, 0_\perp)$
and $p_2/m=\,({\rm cosh} {\gamma}/2, -{\rm sinh} {\gamma}/2, 0_\perp)$
with the rapidity $\gamma$ defined through ${\rm cosh}{\gamma}/2 ={\sqrt s}/2m$.
For $s\gg m^2$ the rapidity gap between the receding scatterers become large
with $\gamma\approx {\rm log} (s/m^2)$. The momentum transfer between the scatterers
is $q=p_1-k_1$, with $q_0\approx q_3\approx t/\sqrt{s}$ and $q_\perp^2= tu/(s-4m^2)\approx -t$.
Hence $q=(0,0,q_\perp)$ with $q^2=-q_\perp^2=t$. Although the partons or dipoles change
their velocities after scattering, this change is small for $s\gg -t$. This is the
kinematical assumption behind the use of the eikonal approximation.

In Euclidean geometry, the kinematics is fixed by noting that
the Lorenz contraction factor translates to
\begin{equation}
{\rm cosh}\,\gamma = \frac 1{\sqrt{1-v^2}} =\frac s{2m^2}-1\rightarrow {\rm cos}\,\theta\,\,.
\label{LO1}
\end{equation}
Scattering at high-energy in Minkowski geometry follows from scattering in
Euclidean geometry by analytically continuing $\theta\rightarrow -i\gamma$
in the regime $\gamma\approx \,{\rm log}\, (s/m^2)\gg 1$~\cite{MEGGIOLARO}. 
It is sufficient to analyze the scattering for 
$p_1/m=\,(1,0,0_\perp)$, $p_2/m=\, 
({\rm cos}\theta\,,-{\rm sin}\theta,\,0_\perp)$, $q=(0,0,q_\perp)$
and $b=(0,0,b_\perp)$.
The Minkowski scattering amplitude at high-energy
can be altogether continued to Euclidean geometry through 
\begin{eqnarray}
&&{\cal T}_{AB,CD} (\theta, q) \approx  
4m^2\,{\rm sin}\,\theta \int d^2b\,\, e^{iq_{\perp}\cdot b}
\nonumber\\&&\times
\la \left({\bf W}(\theta, b) -{\bf 1}\right)_{AC} 
\left( {\bf W}(0,0) -{\bf 1}\right)_{BD}\ra
\label{4X}
\ee
where
\be
{\bf W}(b,\theta)= {\bf P}_c {\rm exp}\left(ig\int_\theta\,d\tau\,
A(b+v\tau)\cdot v\right)
\label{5X}
\ee
with $v=p/m$. The line integral in (\ref{5X}) is over a straight-line
sloped at an angle $\theta$ away from the vertical.

In QCD perturbation theory, different time-ordering contributions to quark-quark 
scattering are shown in Fig.~\ref{FIG1} to order $g^2$. They contribute
to the  T-matrix as ${\cal T}=2{\cal T}_1+2{\cal T}_2$ with 
($T\rightarrow\infty$)~\footnote{The color
factors can be restored trivially.}
\begin{eqnarray}
{\cal T}_1(\theta , b ) =&& \,\frac {g^2}{4\pi^2} 
\,\,\int_0^T\,d\tau_1\int_0^T\,d\tau_2\,\nonumber\\&&
\times\frac {{\rm cos}\,\theta}{(\tau_1-\tau_2\,{\rm cos}\,\theta)^2+\tau^2_2{\rm sin}^2\,\theta +b^2}
\nonumber\\
=&&\frac {\theta}{{\rm tan}\,\theta}\, \frac{g^2}{4\pi^2} \, {\rm log} \left(\frac Tb\right)\,.
\label{PT1}
\end{eqnarray}
and
\begin{eqnarray}
{\cal T}_2(\theta , b ) =&& \frac {g^2}{4\pi^2} 
\,\,\int_0^T\,d\tau_1\int_{-T}^0\,d\tau_2\,\nonumber\\ &&
\times\frac {{\rm cos}\,\theta}{(\tau_1-\tau_2\,{\rm cos}\,\theta)^2+\tau_2^2{\rm sin}^2\,\theta +b^2}
\nonumber\\
=&&\frac {(\pi -\theta)}{{\rm tan}\,\theta}\, \frac{g^2}{4\pi^2} \, {\rm log} \left(\frac Tb\right)\,.
\label{PT2}
\end{eqnarray}
with ${\cal T}_2(\theta , b)=-{\cal T}_1(\pi-\theta, b)$ as expected
from geometry\footnote{The reader may be puzzled by why we are emphasizing
this simple point.  We note that for more involved  multi-gluon
processes this cancellation is spoiled by color factors and
powers of the angle survive in the answer: after the analytic continuation
to Minkowski space these powers become powers of rapidity. They
exponentiate and  produce powers of the collision energy
characteristic of Reggeon behavior (to be described elsewhere).}.
We note that the overall linear dependence in $\theta$ reflects on 
the range of the gluon exchanged in rapidity space caused by our 
ordering in time. This dependence becomes $\theta+ (\pi-\theta)=\pi$ in the sum ${\cal T}$, i.e.
\begin{equation}
{\cal T}(\theta , b) =\frac {g^2}{2\pi^2}\, \frac{\pi}{{\rm tan}\,\theta}
\,{\rm log} \left(\frac Tb\right)\,
\label{PT3}
\end{equation}
as the ordering is unrestricted between $0$ and $\pi$. All gluons between
the spatial distance $b$ and $T$ are also exchanged, hence the infrared 
sensitivity of the quark-quark scattering amplitude in 
perturbation theory. This sensitivity drops from the cross section
(see below).
We note that the order $g^2$ contribution to (\ref{PT3}) is of order $s^0$ after analytical
continuation, in agreement with the general energy-spin assignment for vector exchange. We recall 
that the expected behavior is $s^{J-1}$ for a spin-J exchange.

The contribution of (\ref{PT3}) to ${\cal T}$ follows after integrating over the
impact parameter $b$. The result in Euclidean geometry is
\be
{\cal T} (\theta , q) = &&4m^2\, {\rm sin}\,\theta  \,\int d^2b \,e^{iq\cdot b} \,{\cal T}(\theta , b)
\nonumber\\
=&& -{\rm cos}\,\theta\, \frac{g^2}{2}\, \frac{4m^2}{q^2}\,\int_0^\infty dx\,J_0 (x)\,{\rm log}x \,\,.
\label{PT4}
\ee
which can be translated into Minkowski geometry  by 
analytical continuation through $\theta\rightarrow -i\gamma$
with $q^2=-t$. In both geometries, ${\cal T}$ is purely real and divergent as $t\rightarrow 0$, leading to a differential
cross section of the order of $d\sigma/dt\approx g^4/t^2$ with a
corresponding
divergent Coulomb cross section
$\sigma\approx g^4/(-t_{min})$.
 In perturbation theory, the ${\cal T}$ matrix acquires absorptive parts and turns complex
to higher-order, i.e. ${\cal T}=g^2/t +ig^4/t + ...\,\,$.
The Euclidean perturbative analysis can be carried out to higher orders
as well, in close analogy with analytically continued Feynman 
diagrams~\cite{MEGGIOLARO}.

\begin{figure}[hb]
  \epsfxsize=2.in
  \centerline{\epsffile{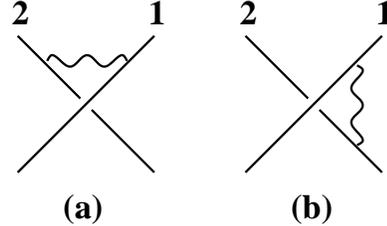}}
  \vspace{0.05in}
  \caption[]{
\label{FIG1}
One-gluon exchange between two receding partons, as discussed in
Eq.(19) (a) and Eq.(20) (b).}
\end{figure}

\subsection{Dipole-Dipole Scattering }\label{DDPT}

We now consider  dipole-dipole scattering 
\be
D_A\,(p_1) + D_B\,(p_2) \rightarrow D_C\,(k_1) +D_D\,(k_2)
\label{3X}
\ee
emphasizing its color degrees of freedom.
For simplicity we will assume both  dipoles to have sizes  $d$, and
(in this section) average over
their orientations. 
For pedagogical reasons, we start with a ``naive'' Euclidean approach
at large impact parameter b,  analogous to the calculation
of the dipole-dipole potential above. This would be shown to 
lead to an incorrect answer for the high energy scattering amplitude.
The reason will be given below along with the correct answer.

We will assume that the impact parameter $b$ is large in comparison
to the typical time characteristic of the Coulomb interaction inside
the dipole, i.e. $b\gg \tau_0\approx d/g^2$. In the elastic 
dipole-dipole amplitude the dipoles remain color-neutral, and may
argue that the leading order is 2-gluon dominated. In analogy to the
potential, one may rely on the Coulomb interaction inside the dipole
to write the dipole-dipole effective vertex in the form
\begin{equation}
S_{eff}= {\alpha_E} \,\int_{-\infty}^{\infty} \,d\tau\, 
\dot{x}_{\mu} \dot{x}_{\nu}\,F_{\mu\alpha}^a\,F_{\nu\alpha}^a (x)
\label{DD1}
\end{equation}
where the electric polarizability $\alpha_E\approx (gd)^2/{\cal E}$ 
with ${\cal E}\approx g^2/d$ its Rydberg energy~\cite{PESKIN}. 
(Higher order operators are suppressed by powers of the dipole size $d$.)

In leading order in the dipole size, the scattering amplitude
then reduces to
\be
{\cal T} (\theta, b)\approx &&  \,{\alpha_E}^2\,
\dot{x}_{1\mu}\dot{x}_{1\nu}\dot{x}_{2\lambda}\dot{x}_{2\sigma}\nonumber\\
&&\times
\int_{-\infty}^\infty \, d\tau_1\,d\tau_2\,
\la F^a_{\mu\alpha}F^a_{\nu\alpha} (x_1) \,
 F^b_{\lambda\beta}F^b_{\sigma\beta} (x_2) \ra\,\,,
\label{DD2}
\ee
with $x_1=v_1\tau_1$ and $x_2=v_2\tau_2 +b$.
The last expectation value can be unwound using free
field theory to obtain 
\begin{equation}
{\cal T} (\theta, b)
\approx \frac {(N_c^2-1)}{\pi^3}\,
\frac{\alpha_E^2}{b^6}\,
\left(\frac {11}{25}\frac 1{{\rm sin}\,\theta} + \frac 85 \frac{{\rm cos}^2\theta}{{\rm sin}\,\theta}\right)\,\,.
\label{DD3}
\end{equation}
We note that the result
(\ref{DD3}) diverges as $\theta\rightarrow 0$. For the case $\theta=0$, we obtain
the Casimir-Polder-type amplitude 
\begin{equation}
{\cal T} (0,b)\approx \frac {(N_c^2-1)}{\pi^3}\,
\frac{T\alpha_E^2}{b^7}\,\frac {23}{8}\,
\label{DD4}
\end{equation}
with $T\rightarrow\infty$, which differs from the $\theta\neq 0$ by the occurrence 
of the infrared sensitive factor $T/b$.

The analytical continuation of (\ref{DD3}) to Minkowski space 
shows that the first contribution is of order $1/s$, while the second
contribution is of order $s$.  This implies that the total cross section
is unbound, i.e. $\sigma \sim s$,
 which is clearly incorrect. Indeed, on physical grounds
the total cross section should be constant at large $s$. 
In Minkowski space it is easy to understand what went wrong. The
electric field of a boosted dipole looks like a Lorenz contracted 
disk with a very small longitudinal width $b/{\rm ch\,}y\ll b$. 
Clearly, at high energy the interaction time of two dipoles is of
this order of magnitude, which is much shorter than the Coulomb time
$\tau_0$. During this short time, the color rotation induced by the
Coulomb interaction can be ignored. Therefore, the use of (\ref{DD1})
in the form of a local 2-gluon exchange is incorrect~\footnote{Note that the result (\ref{DD4}) is not based on this approximation, and therefore
is still valid.}. This point
is actually missed in the Euclidean formulation as the Lorenz factor
is ${\rm cos}\,\theta\sim 1$. Although any particular integral can be analytically
continued from Euclidean to Minkowski space, kinematical approximations 
can only be inferred from the Minkowski domain where all parameters
have their physical values. This will be understood throughout.

So ignoring the Coulomb interaction and
 using the eikonal approximation,
LSZ reduction and the analytical continuation discussed above, 
we can  write
the dipole-dipole scattering amplitude ${\cal T}$ in Euclidean
geometry similarly to (\ref{4X}) with 
\be
{\bf W} (\theta, b)= \frac 1{N_c} {\rm Tr} \left( {\bf P}_c {\rm exp}\left(ig\int_{{\cal C}_{\theta}}\,d\tau\,
A(x)\cdot v\right)\right)
\label{5XX}
\ee
where $x$ is an element of ${\cal C}_{\theta}$. In Euclidean geometry
${\cal C}_{\theta}$  is a closed rectangular loop of width $d$, that is
slopped at an angle $\theta$ with respect to the vertical direction.
To leading order in the dipole-interaction, ${\cal T}$ can be
assessed by expanding each Wilson-line (\ref{5XX}) in powers of $g$, 
and treating the resulting 2-gluon correlations perturbatively.
The result is
\be
{\cal T} (\theta , b) \approx \frac {N_c^2-1}{N_c^2}\,\frac {(gd)^4}{32\pi^2}\,
\frac{{\rm cotan}^2\,\theta}{b^4}
\label{6X}
\ee
for two identical dipoles $d_1=d_2=d$ with polarizations along 
the impact parameter $b$. For small size dipoles, (\ref{6X}) 
is the dominant contribution to the scattering amplitude.
The analytical continuation shows
that ${\rm cotan}\,\theta\rightarrow -i$, leading to a finite total
cross section as expected.

\section{ Instanton effects on the potentials}\label{INST}

\subsection{Generalities}

Instantons are self-dual solutions to the classical
Yang-Mills equations in vacuum originally discovered in ref. \cite{BPST}.
They are classical paths describing tunneling between topologically
inequivalent vacua of the gauge theory.  In QCD, instantons
were argued to be  responsible for observable  phenomena such as
the resolution of the U(1) problem (large $\eta'$ mass)~\cite{tHooft} and 
the spontaneous breaking of chiral symmetry~\cite{Shu_82,DP_86}.  
The interacting  instanton liquid model (IILM) has been 
shown to reproduce multiple correlation functions, including 
hadronic spectra and coupling constants (for a review see \cite{SS_98}).

Instantons are also commonly  used in other gauge theories,
especially in supersymmetric gauge theories where supersymmetry
makes their effects dominant in the non-perturbative regime. 
Indeed, some 
exact results (such as the effective low energy Lagrangian for 
$N$=2 supersymmetric theories derived by Seiberg and Witten
and also the AdS/CFT correspondence suggested by Maldacena
~\footnote{In fact, the 5-dimensional anti-de Sitter space emerges
from the space of the instanton collective coordinates
(the center position and size $d^4z d\rho/\rho^5$) which  will be 
extensively used for averaging  below.}
for the N=4 super-conformal theory) can be exactly reproduced using 
exclusively the instanton calculus developed in~\cite{Mattis_etal}.

For the purpose of this paper the topology of instantons is not 
important: heavy quarks do not interact with fermionic zero modes,
and high energy quarks for all purposes behave as heavy quarks.
What is important instead is the following technical point:
in the instanton field the path-ordered exponents can be 
evaluated $analytically$, since the color phase rotations take
place around the same axis for a fixed path
(the instanton is a hedgehog in color-space). The self-duality
of the instanton field will also have an effect on some of our
results. Once a path-ordered exponent
is evaluated in the one-instanton field, the vacuum averages 
follow through the instanton ensemble average representing the
QCD vacuum (dilute phase). This includes averaging over the
instanton center-position $z_\mu$ and size $\rho$.
Specifically, we will use the measure 
\be
\label{dist0}
dn = d\rho\, d^4 z \,{ D(\rho) \over \rho^5}\ee
for both instantons and anti-instantons.
The  integral over $z$ can be sometimes carried out analytically, 
but most of the times will be done numerically.
The understanding of the instanton size distribution $D(\rho)$
remains an open problem. Naive 
semi-classical results suggest~\cite{tHooft}:
\begin{eqnarray}
 D_0(\rho)\approx &&C_{N_c}({8\pi^2 \over
  g^2(\rho)})^{2 N_c}{\rm exp}(-{8\pi^2 \over
  g^2(\rho)})\nonumber\\ \approx &&  (\rho \Lambda)^{(11/3)N_c-(2/3)N_f} 
\ee
where $C_{N_c}$ is a constant depending on the number of colors $N_c$.
We have used the asymptotic freedom formula in the exponent to
show that this density dramatically grows with the instanton size $\rho$. 
However in the true QCD vacuum instantons and antiinstantons
interact with each other and othe quantum fields, so that 
the real function $D(\rho)$ deviates from the semiclassical
one for large sizes. 

For qualitative estimates we will often use parameters
of the instanton liquid model \cite{Shu_82}, which assumes that all
instantons have the same size
\be dn(\rho) =n_0 \,d^4z\, d\rho\, \delta(\rho-\rho_0) \ee
where $n_0$ is the total instanton (plus
anti-instanton) with a typical radius $\rho_0$, i.e.
\be
n_0\approx 1 \,\, fm^4;\,\,\, \rho_0\approx 1/3\, fm 
\ee
These values
were deduced from phenomenological data extracted from
the QCD sum rules, 
the topological succeptibility 
and the chiral condensate long before direct lattice data
became available. In Fig.~\ref{rhodist} we show a sample of such lattice
measurements, together with the parameterization
for the instanton suppression suggested in
\cite{Shu_sizes}.  Specifically,

\begin{figure}[hb]
  \epsfxsize=3.in
  \vspace{-.1in}
  \centerline{\epsffile{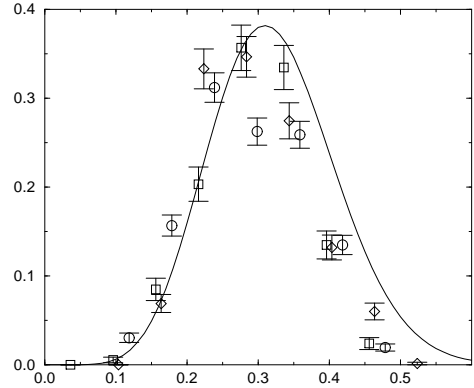}}
  \vspace{-.05in}
  \caption[]{
\label{rhodist}
(a) The instanton density $dn/d\rho d^4z$, [fm$^{-5}$] versus its size
 $\rho$ [fm]. (b) The combination  $\rho^{-6} dn/d\rho d^4z$, in which
 the main one-loop behavior drops out for $N_c=3,N_f=0$.
 The points are from the lattice work \protect\cite{anna},
for this theory, with 
$\beta$=5.85 (diamonds), 6.0 (squares) and 6.1 (circles). Their
comparison should demonstrate that results are
rather lattice-independent.
The line corresponds to \protect\cite{Shu_sizes}.
  }
\end{figure}
\be 
dn(\rho)= dn_0(\rho) e^{-2\pi\sigma\rho^2}
\ee
can be used for averaging in any integral over the instanton density.
Typically, the string tension $\sigma\approx (0.440 \, MeV)^2$, so
that
\begin{eqnarray}
&& <\rho^2>\approx (0.28 \,fm )^2;\nonumber\\
&&  <\rho^4>\approx (0.31
\,fm)^4;\nonumber\\
&&<\rho^5>\approx (0.32 \,fm)^5\,\,,
\end{eqnarray}
which shows that the difference between the realistic averages 
and simple powers of $\rho_0$ are relatively small. We will ignore 
these differences below.

In the analysis to follow, the parameters capturing the instanton
physics will appear as two dimensionless quantities:
(i) a small {\em diluteness} parameter and (ii) a large  {\em  action} of
an instanton (per $\hbar$):
\be
n_0 \rho_0^4\approx \left({1\over 3}\right)^4 \,\, \qquad
\,\,
S_0={8\pi^2 \over g^2(\rho_0)}\approx (10-15)
\ee
The small factor is a penalty for finding the instanton, and
the large factor is a classical enhancement relative to perturbation
theory.  Their interplay would cause particular effects to be 
parametrically large or small.

\subsection{Static  quarks }

At the one instanton level, the various potentials for a static quark-antiquark
potential have been assessed  long ago  \cite{CDGWZ}, including the spin-dependent part.
We will briefly review this assessment for completeness. We recall that the 
various components of the potential follow from the rectangular $T\times R$
Wilson loop
\be
V(R)=-\frac 1T \,\lim_{T\rightarrow\infty}{\rm ln} \,<{\bf W}(T,R)>
\ee
evaluated in a  classical instanton field, after averaging over the instanton
position. In the Wilson loop, the path-ordered exponents $P exp(ig \int A_\mu dx_\mu)$
can be evaluated analytically as the instanton locks the color orientation to space.
Indeed, the  static potentials involve $A_0^a\sim \eta^a_{0,\nu} (x-z)_\nu
\sim  (x-z)^a$ where $(\vec x-\vec z)$ refers
to the distance between the quark position and the 3-d coordinate of the
 instanton
center\footnote{The time position
of the instanton $z_4$ is  irrelevant.}.
The resulting color rotation angle $\alpha$                    
\cite{CDGWZ} and the unit vector $n^a$ around which the rotation takes place
are defined through
\be \label{line}
{\bf W}=&&{\rm exp}\left(-i\pi {\tau^a (z_a-r_a) \over ((r_a-z_a)^2+\rho^2)^{1/2}}\right)\nonumber\\=&&
{\rm exp}(-i\pi \tau^a n_a \alpha)\,.
\ee
 If all relevant distances are comparable, $|r_a-z_a|\sim \rho_0$,
the rotation angle is O(1), showing that the expansion in field
strength is in general not justified. For a small-size
dipole, the potential is  small $V(R\rightarrow 0)\approx R^2$,
since the path-ordered lines in ${\bf W}$ are close enough to cause
partial cancellation.
However when $R\approx \rho_0$, and both path-ordered lines happen to be 
on the  opposite sides of the instanton center, the color rotations on 
both lines adds up and the  potential becomes roughly linear in R and
more
sizable. Finally, when the dipole is too large, the potential
saturates\footnote{In ref.\protect\cite{Shu_toward} one of us has
  noticed that this behavior is surprisingly similar to that
  experimentally observed in deep inelastic scattering, if $Q^2$
  dependence of structure functions is treated as
dependence of the cross section on the   dipole size. }

The quark-antiquark potential calculated in
\cite{CDGWZ} can be expressed as 
\be \label{def_F} V(R)=  \int dn(\rho) \rho^3 F(R/\rho) \ee
where 
the dimensionless function F is defined as
\be
F =\int {d^3z\over N_c \rho^3} {\rm Tr} (1-{\bf W}_1 {\bf W}_2^\dagger)\,\,.
\ee
The trace-part of the integrand is 
\be
2\left(1-{\rm cos}\,\alpha_1\,{\rm cos}\,\alpha_2-\,\vec n_1\cdot\vec n_2\,\,
{\rm sin}\,\alpha_1\,{\rm sin}\,\alpha_2\right)\,\,.
 \ee
where the angles $\alpha_i$ and vectors $n_i^a$
are defined  in (\ref{line}). 
This function is shown in Fig.~\ref{fig_instpot}~a. 
In order to emphasize the small-R ``dipole limit'' $V(R)\sim R^2$
 (to be important
for what follows), we have also plotted the ratio of this function
to
its dipole limit in Fig.~\ref{fig_instpot}~b. One can see 
 that the dipole approximation has an unexpectedly large
range of applicability: this ratio does not change appreciably
(less than 25\%)  till
$R\approx\rho_0$. One may expect similar accuracy of the dipole
approximation in other applications to be discussed.

\begin{figure}[t]
\includegraphics[width=3.in,angle=-90]{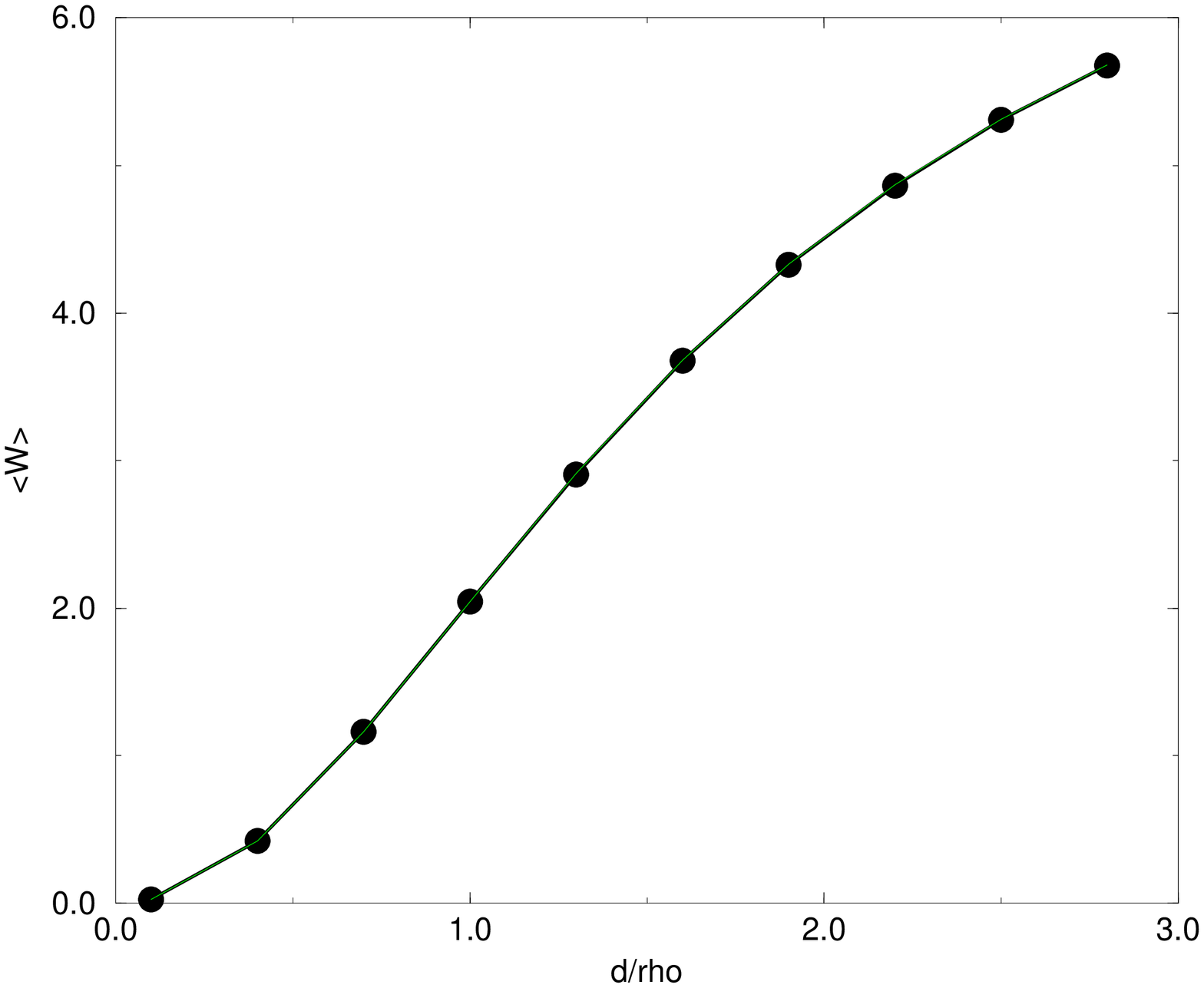}
\vskip -0.2in
\includegraphics[width=3.in,angle=-90]{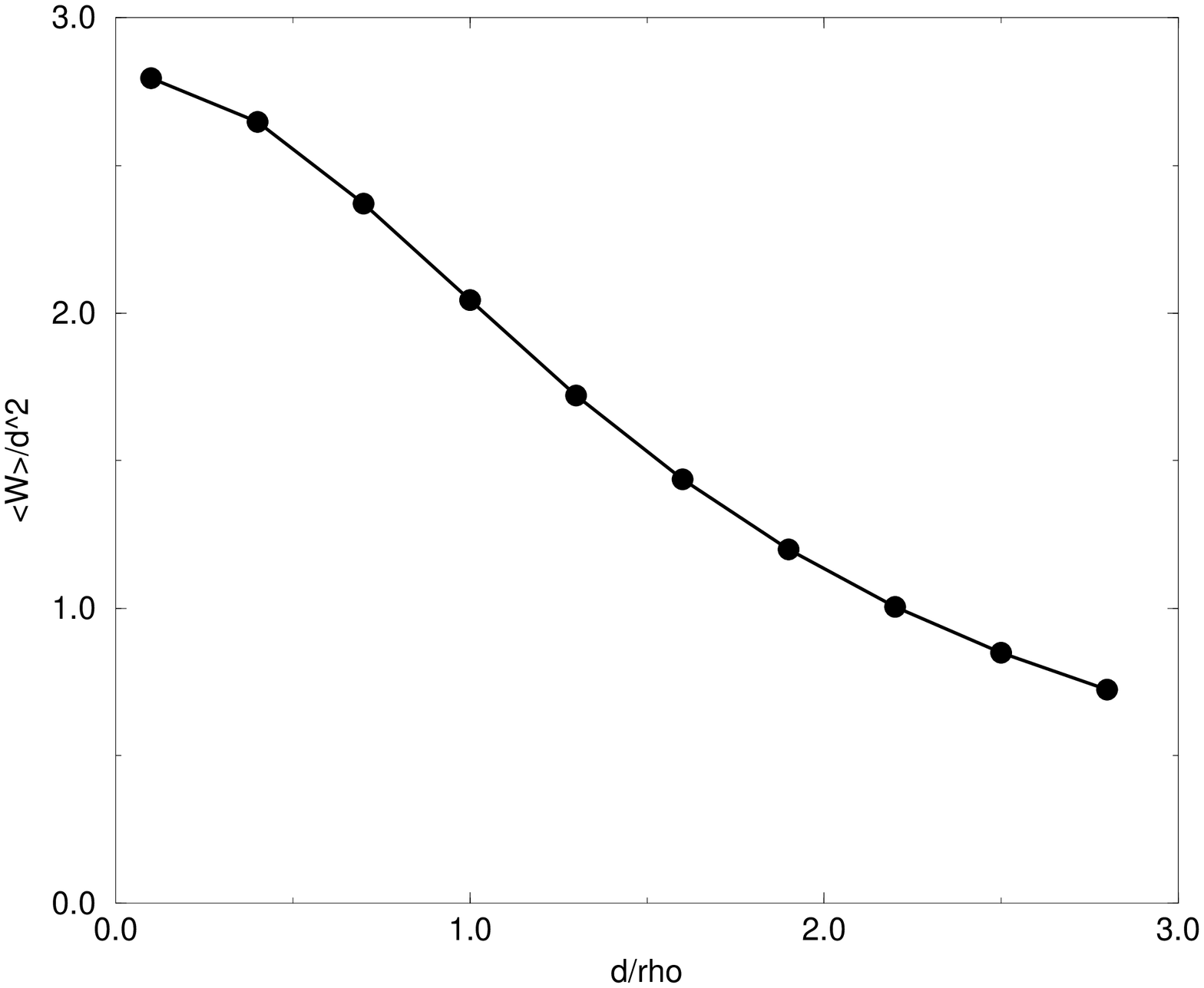}
\caption[]{
 \label{fig_instpot}
(a) F as defined in (\ref{def_F}) as a function of the quark-antiquark
distance R measured in units of the instanton size $\rho_0$; (b)
the rescaled function $F(R/\rho_0)/R^2$ to exhibit 
the accuracy of the dipole approximation at small R.
}
\end{figure}

For large R the potential \cite{CDGWZ}
goes to a constant plus a Coulomb term 
\be 
V(R\rightarrow \infty)=37 \int {d\rho \over \rho^2} D(\rho)- {4\pi^3 \over 3 R}{d\rho \over \rho} D(\rho)+...
\ee
which can be interpreted as the instanton contribution to
the $mass$ and $charge$ renormalization, respectively. It is instructive
to compare the magnitude of the latter 
 to the perturbative potential, through 
\be \label{eq_chargerenorm} 
{V_{inst} \over V_{pert}}= {\pi^2 \over 2}  (n_0 \rho_0^4) ({8\pi^2
  \over g^2(\rho_0)}) \,\,,  \ee
with $V_{pert}=4\alpha_s/3R$.
The ratio is the product of the {\em diluteness parameter}~\footnote{
The coefficient in front of $\pi^2\rho^4/2$ happens to be the volume
of a 4-sphere.} (the
fraction of space-time occupied by instantons) times the  {\em classical
enhancement} through the instanton action (per $\hbar$).
Using the phenomenological parameters discussed above,  we
observe that the diluteness is compensated by the classical
enhancement, so that the instanton corrections 
at $R\approx \rho_0$ are actually comparable to the perturbative
Coulomb effect.

However, instantons are not the only non-perturbative effects
contributing to the static quark-antiquark potential. At large $R$
confinement in the form of a QCD string with $V_{conf}\approx \sigma R$
dominates. In fact, already for $R\approx\rho_0\sim 0.3 fm$ confinement
is dominant, with the instanton-induced potential accounting for
only 10-15\%~\footnote{The claim
made in \cite{jap_conf}, that instanton effects account for the
confining potential is incorrect.}. For a detailed study of
these issues at the  multi-instanton level, one can consult
refs \cite{pot_numeric} for a numerical analysis and \cite{DP_pot} for analytical results.

\subsection{Static dipoles}

Unlike the quark-antiquark potential, the dipole-dipole potential
is insensitive to confinement, and the instanton-induced
interaction may be easier to identify. In the latter case, we will
consider two cases where the characteristic time within the dipole
is either  (i) $short$ $\tau_0\sim d/g^2\ll \rho$ or (ii) $long$ 
$\tau_0\gg \rho_0 $ in comparison to the instanton size. These two
cases translate to a magnitude of the dipole field $A_0\sim
g/d$ which is large (i) or small (ii) in comparison 
to that of the instanton field $A_\mu\sim 1/g\rho$. 

 In the  case (i) the static potential can be written  in terms of
 the polarizabilities, and the correlator of gluo-electric
 fields
\be
V(R) = {\alpha_1 \alpha_2 } \int d\tau <\vec E^2(\tau, R)\vec E^2(0,0)> 
\ee
 This field strength  correlator can be
evaluated by substituting the expression for the instanton field
\be
\vec E^2(x)=\vec B^2(x)=  {96\rho^4\over g^2}{1 \over ((x-z)^2+\rho^2)^4}
\ee
The averaging of the correlator over  the location of the instanton position $z$ 
 can carried out analytically \cite{SS_95}
\begin{eqnarray}
<(gG_{\mu\nu}^a(x))^2(gG_{\mu\nu}^a(0))^2>=\nonumber\\{384 g^4 \over \pi^4 x^8}+ (n_0\rho_0^4)
\,\,\Pi_{inst}(x/\rho)/\rho^8 \,\,,
\end{eqnarray}
where the last term was added to account for the perturbative contribution. The dimensionless
function describing the instanton contribution is

\begin{figure}[ht]
\includegraphics[width=3.6in,angle=-90]{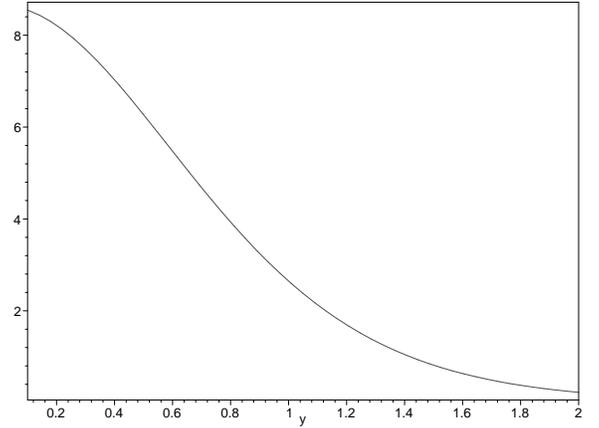}
\vskip -0.6in
\includegraphics[width=3.6in,angle=-90]{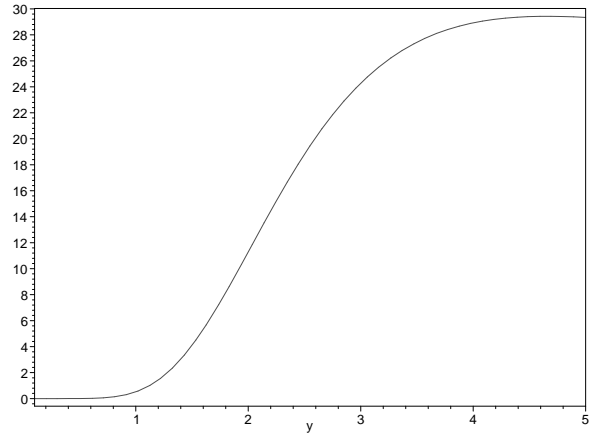}
\vskip -0.6in
\caption[]{
 \label{fig_maple}
(a) Field strength correlator $\Pi_{inst}$ as defined in 
(\ref{PIXX}) versus the distance in units of the instanton size
$x/\rho_0$. (b) Ratio of the instanton-induced term in the correlator
  to the perturbative one,
 versus the distance. 
}
\end{figure}

\be
\Pi_{inst}(y)=&& \frac {12288\pi^{2}}{y^{6}\,(y^{2} + 4)^{5}}\nonumber\\&&\times
 (  y^{8} + 28\,y^{6} - 94\,y^{4} - 160\,y^{2} - 120  \nonumber\\&&
   + \frac {240}{y\,\sqrt{y^{2} + 4}} \nonumber\\
&&\times (y^{6} + 2\,y^{4} + 3\,y^{2} + 2)\,{\rm arcsinh}(y/2) )\,\,.
\label{PIXX}
\ee
Its behavior is shown in Fig.\ref{fig_maple}a.
Its ratio to the perturbative contribution 
to the same correlator (for $g=2$ or $\alpha_s=0.32$) is shown in
fig.\ref{fig_maple}b. As expected, it is small at small distances
$x\ll \rho_0$. At large distances,  
the instanton-induced contribution has the same behavior $\Pi_{inst} \approx
1/R^{8}$, as
the perturbative one. Furthermore, the ratio of the two
is about 30, much more  than the
 ``instanton-induced charge renormalization'' (\ref{eq_chargerenorm}) we
discussed
in the preceding subsection. About the same is found in the 
potentials themselves
(the correlator integrated over the time difference)
as shown in  Fig.\,~\ref{fig_maple}~d. 
The perturbative behavior is
dominated by $two$ gluons rather than one, and therefore the instanton
effect occurs with a classical enhancement $squared$:
\be {V_{inst} \over V_{pert}} \sim (n_0 \rho_0^4)\, \left({8\pi^2 \over g^2(\rho_0)}\right)^2    \ee 
This feature
 implies that instanton effects
are much more important for dipole-dipole interactions at $R\approx
\rho_0\approx 0.3$ fm than the perturbative Casimir-Polder effects. 
We will argue below that this is generic for all processes demanding
multi-gluon exchanges, and that instanton-induced processes can become
dominant in this case.

In the case ii), the dipoles can
be considered quasi-static in time, $\tau_0\sim d/g^2\gg\rho_0$, and 
the  time evolution of the color degrees of freedom due to
the  Coulomb
interaction can be ignored. In other words,  the  dynamics 
 is driven entirely by the instanton field.
The potential between two dipoles is now
\be
 \label{def_dd} V_{dd}(R)=  \int dn(\rho) \rho^3 F_{dd}(R/\rho) \ee
with
\be
F_{dd} =\int {d^3z\over N_c \rho^3} (1-{\rm Tr}{\bf W}_1 {\rm Tr} {\bf W}_2) \ee
Here {\bf W} are rectangular Wilson $loops$ for each dipole, traced
separately.  Averaging over the instanton position can be done numerically.
The results are shown in Fig.\ref{fig_dd}. The outcome
is  proportional to $d_1^2 d_2^2$ (dipole-moments)
rather than $\alpha_1\,\alpha_2$ (electric polarizabilities),
when $d$ is reasonably small in comparison to $\rho_0$.
 The large distance
potential is few \% that of  $V(R)\approx d_1^2 d_2^2 \rho_0^2 /R^7$.
 Note that it is larger
than the perturbative one since $\rho_0^2$ is assumed to be much larger than
 $d_1d_2$, but both answers have the same (zeroth) power of g.

In general the dipole-dipole potential cannot be
approximated by  the correlator of scalars $E^2$, as can be
checked through its dependence on the relative orientation of the dipoles.
Even in the dipole (quadratic) approximations for sufficiently small dipoles
($d_i\ll\rho$) one can define  4 invariant functions for the
dipole-dipole interaction
\begin{eqnarray}
 \label{invariants}
V(R)=&&d_1^i d_1^j d_2^l d_2^m\,\,(
A(R)\delta_{ij}\delta_{lm}+\nonumber\\&&
\frac 12 B(R)(n^i n^j \delta_{lm}+n^l n^m \delta_{ij})
+\nonumber\\&&
 C(R)n^i n^l \delta_{jm}+D(R) n^i n^j n^l n^m \,\,)
\end{eqnarray}
The first function A(R) 
accounts for the spin-zero gluonic operator $\vec E^2$
discussed at the beginning of this subsection.
 However, as one can see from fig(\ref{fig_dd}),
other functions also contribute. In (a) we compare the xx orientation
(or A+B+C+D) with the xy one (or A+B/2) and see a clear difference.
In (b) we note the dependence on the rotation angle for one of the
dipoles, which shows a clear ${\rm cos}^2\,\theta$ behavior expected from the
expression above.

\begin{figure}[ht]
\includegraphics[width=3.in]{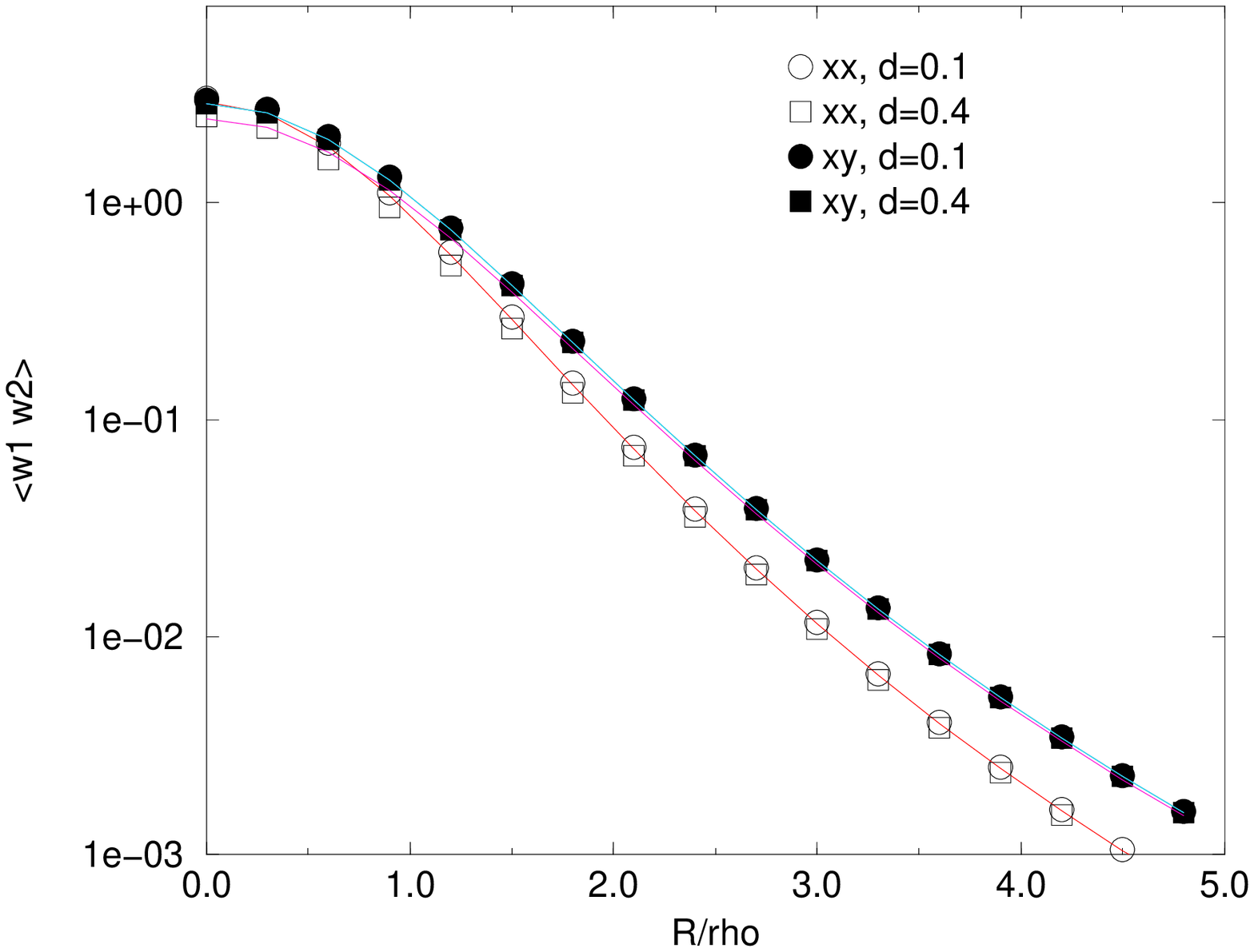}
\vskip -0.6in
\includegraphics[width=2.5in,angle=-90]{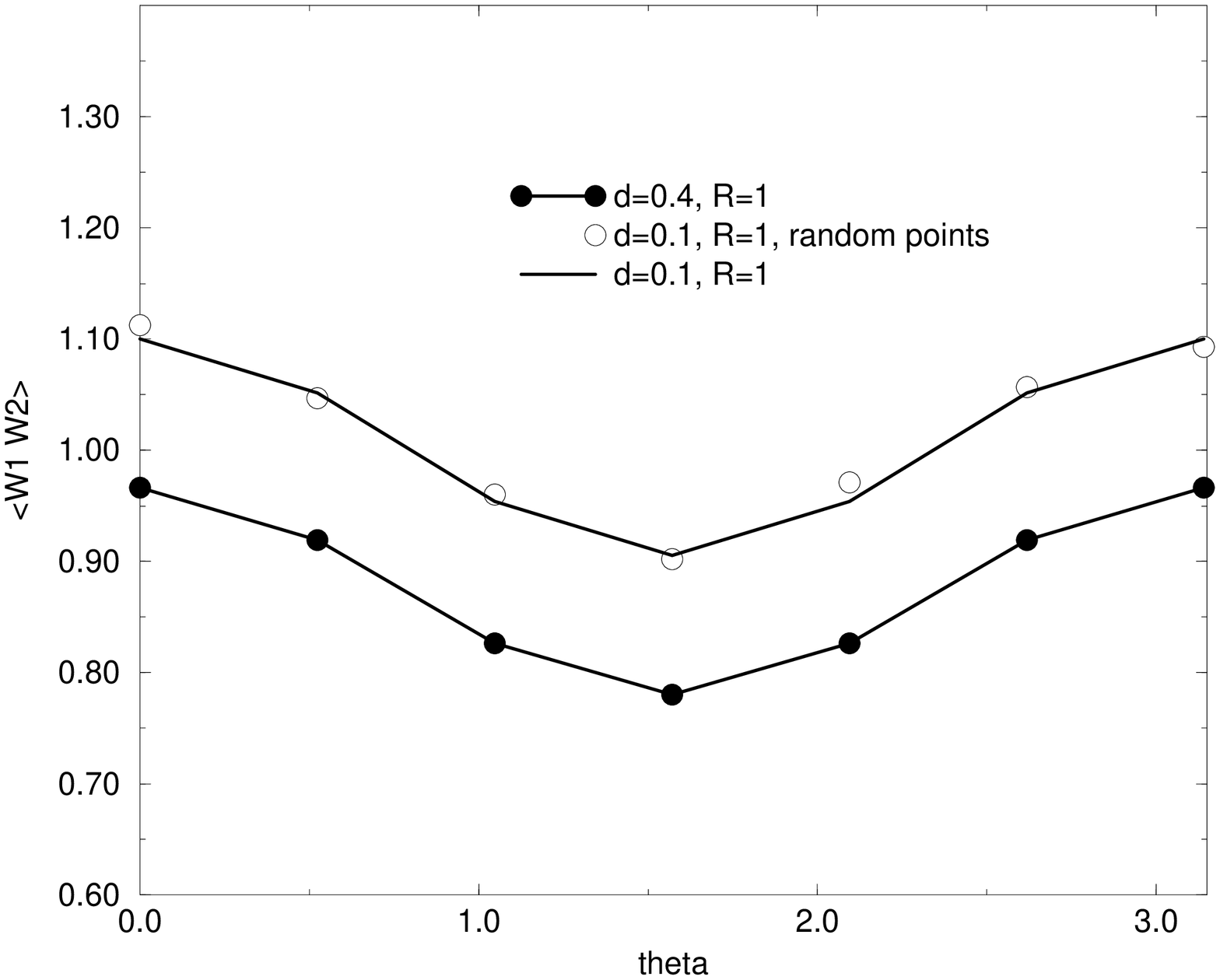}
\caption[]{
 \label{fig_dd}
(a)Two correlated  Wilson loops as a function of the distance R between
their centers, divided by $d_1^2d_2^2$ for two
dipole sizes, $d_1=d_2=0.1$ (circles) and 0.4 $\rho_0$ (squares).
The agreement between the points show that the dipole scaling holds well for such sizes.
Also two dipole orientations are shown. 
The open points are for both dipoles  oriented in the x
direction
(the same direction as the distance between them R) while the closed points for
the xy
orientation. The disagreement between those means that the dipole-dipole forces
depend on the orientations. Further details on the orientation dependence are
shown in Fig.(b). $F_{dd}(R=\rho_0)$ shows the dependence on the
orientation angle of one dipole in the x-y plane. The solid points and curves
are for a 4-d lattice-type integration over the instanton center, and the open
points are for the alternative Monte-Carlo integration: their spread from
the curve shows the magnitude of the uncertainties involved.}
\end{figure}

\section{One-Instanton effect on scattering}

\subsection{Quark-Quark scattering}

Our first step now is the generalization of (\ref{line}) to an arbitrary
orientation $\theta$  of the Wilson  line. The analytical continuation
to Minkowski space follows from $\theta\rightarrow iy$ with $y$ identified
as the rapidity difference between the receding partons. The untraced
and tilted Wilson line in the one-instanton background reads
\begin{equation}
{\bf W} (\theta, b) = {\rm cos}\,\alpha -i\tau\cdot\hat{n}\,{\rm sin}\,\alpha
\label{QQI1}
\end{equation}
where
\begin{equation}
n^a ={\bf R}^{ab}\,\eta^b_{\mu\nu}\,\dot{x}_\mu (z-b)_\nu ={\bf R}^{ab}\,{\bf n}^b
\label{QQI2}
\end{equation}
and $\alpha=\pi\gamma/\sqrt{\gamma^2+\rho^2}$ with
\begin{eqnarray}
\gamma^2=&&n\cdot n={\bf n}\cdot{\bf n}\nonumber\\=&&
(z_4{\rm sin}\theta - z_3 {\rm cos}\theta )^2 + (b-z_\perp)^2\,\,.
\label{QQI3}
\end{eqnarray}
The one-instanton contribution to the
untraced QQ-scattering amplitude
follows from the following correlator
\be
&&\left<{\bf W}_{AC} (\theta , b )
{\bf W}_{BD} (0, 0  )\right> \approx\,n_0\, \int\,d^4z\,\nonumber\\
&&\times({\rm cos}\alpha\,{\rm cos}\underline{\alpha} 
\,\,{\bf 1}_{AC}\,{\bf 1}_{BD}\nonumber\\&&
- \frac 1{N_c^2-1}\hat{\bf n}\cdot\underline{\hat{\bf n}}\,
{\rm sin}\alpha\,{\rm sin}\underline\alpha\, 
\,\,(\tau^a)_{AC}\, (\tau^a)_{BD})\,\, ,
\label{QQI4}
\ee
where the (under) bar notation means the same as the corresponding
un-bar one with $\theta=0$ and $b=0$. 

 Furthermore, 
\begin{eqnarray}
&&\left<\frac 1{N_c}\,{\rm Tr} \left( {\bf W} (\theta , b) \,{\bf W} (0,0) \right) \right>=\nonumber\\&&
\frac {2n_0}{N_c}\int\,d^4z
\left({\rm cos}\,\alpha\,{\rm cos}\,\underline{\alpha} -\hat{\bf n}\cdot\underline{\hat{\bf n}}\,
{\rm sin}\,\alpha\,{\rm sin}\,\underline\alpha\right)\, .
\label{QQX4}
\end{eqnarray}
The integrand in (\ref{QQX4}) can be simplified by changing
variable $(z_4\,{\rm sin}\,\theta -z_3\,{\rm cos}\,\theta )\rightarrow z_4$
and dropping the terms that vanish under the z-integration. Hence
\begin{eqnarray}
&&\left<\frac 1{N_c}\,{\rm Tr} \left( {\bf W} (\theta , b) \,{\bf W} (0,0) \right) \right>=
\frac {2n_0}{N_c}\,\int\,d^4z\nonumber\\
&&\left(\frac 1{{\rm sin}\theta}\,
{\rm cos}\,\tilde\alpha\,{\rm cos}\,\underline{\tilde\alpha} -
\frac 1{{\rm tan}\,\theta} \,{\rm sin}\,\tilde\alpha\,{\rm sin}\,\tilde{\underline{\alpha}}
\, \frac {z_\perp^2-z_\perp\cdot  b}{\tilde\gamma\,\underline{\tilde\gamma}}\right)\,\,.
\label{QQX5}
\end{eqnarray}
The tilde parameters follow from the un-tilde ones by setting $\theta=\pi/2$.
We note that $\underline{\tilde\gamma}=\underline\gamma=|\vec z|$. 
After analytical continuation, the first term produces the elastic
amplitude which decays as $1/s$ with the energy. The second term
corresponds to the color-changing amplitude. It is of order $s^0$
and dominates at high energy. Specifically
\begin{eqnarray}
&&\left<\frac 1{N_c}\,{\rm Tr} \left( {\bf W} (\theta , b) \,{\bf W} (0,0) \right) \right>=\nonumber\\&&
\frac {2n_0}{N_c}\,\,
\left(\frac 1{{\rm sin}\theta}\,\, F_{cc} (b/\rho_0) 
-\frac 1{{\rm tan}\,\theta} \,\,F_{ss} (b/\rho_0)\right)\,\,.
\label{QQX6}
\end{eqnarray}
We show in Fig.\ref{fig_ss} the numerical behavior 
of the two contributions in (\ref{QQX6}). Note that
the second function (which
describes color-inelastic collisions and
 survives in the high energy limit) changes
sign, before  decreasing as a power law to zero at large b.

\begin{figure}[t]
\vskip -0.04in
\includegraphics[width=3.in,angle=-90]{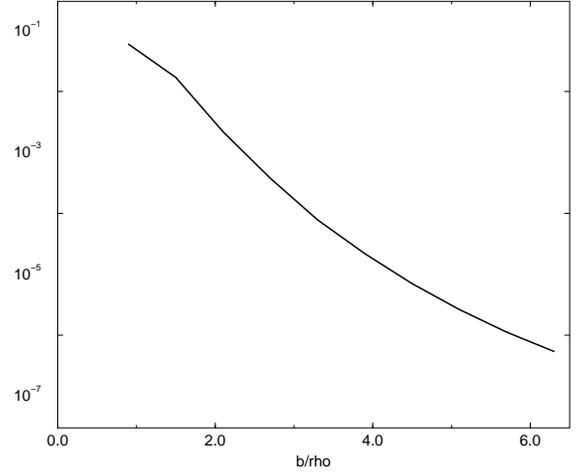}
\vskip -0.04in
\includegraphics[width=3.in,angle=-90]{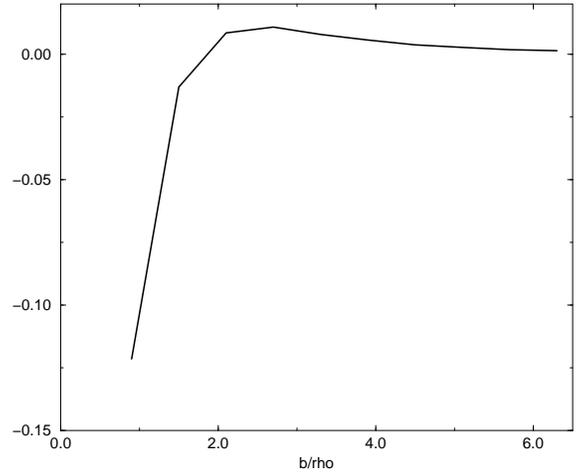}
\caption[]{
 \label{fig_ss}
(a) and (b) show the two functions
$n_0 F_{cc}$ and $n_0 F_{ss}$
 defined in (\ref{QQX6}), 
versus the impact parameter b (in units of the instanton radius).
}
\end{figure}

\subsection{Dipole-Dipole and Multi-parton Scattering}

  One can directly generalize the calculation of the quark-quark scattering amplitude
  to
that of any number of partons.  For that, we assume that they all move with high
energy
in some reference frame and opposite direction: in Euclidean space those
would propagate along two directions, with  parton numbers
$N_1$ and $N_2$ respectively. Any one of them, passing through the instanton
field,
is rotated in color space by a different angle $\alpha_i$ around
a different
axis $\vec n_i$, depending on the shortest
distance between its path and
the
instanton center. Integration over all possible color
orientations of the instanton leads then to global color conservation.

  Before discussing specific cases in details, let us make a general
  qualitative statement about such processes. We have found in the
  previous
section that (the color-changing) quark-quark instanton-induced
scattering has a finite high energy
limit. For  perturbative n-gluon exchange a factor of $\alpha_s^n$
is paid, while for an instanton mediated scattering a factor
of $n_0\rho_0^4$ is paid (the price to find the instanton at the
right place), no matter how many
partons participate. Since the instanton vacuum is dilute, the one-gluon
mediated process dominates the instanton one. However, the situation 
dramatically changes for two or more gluon exchanges:
 the instanton-induced amplitude is about
the same for any number of partons,
 provided that all of them pass at a distance
$\approx \rho_0$ from the instanton center.

Now, consider a dipole configuration of size $d$ chosen in the transverse
plane of a $\overline q q$ located on a straight-line sloped at an angle
$\theta$ in Euclidean space. Let $AA$ be the initial color of the dipole
and $CD$ its final color. The Wilson loop with open color  for the dipole 
configuration in the one-instanton background is
\be
&&{\cal W}_{AA}^{CD} (\theta , b) =
{\rm cos}\,\alpha_-\, {\rm cos}\, \alpha_+\, {\bf 1}_{CD}\nonumber\\
&&+i{\rm cos}\,\alpha_-\,{\rm sin}\,\alpha_+\,{\bf R}^{ab}\,\hat{\bf n}_+^b\,(\tau^a)_{DC}\nonumber\\
&&-i{\rm sin}\,\alpha_-\,{\rm cos}\,\alpha_+\,{\bf R}^{ab}\,\hat{\bf n}_-^b\,(\tau^a)_{DC}\nonumber\\
&&+{\rm sin}\,\alpha_-\,{\rm sin}\,\alpha_+\,{\bf R}^{ab}\,{\bf R}^{cd}\hat{\bf n}_-^b\, \hat{\bf n}_+^d\,
(\tau^c\tau^a)_{DC}\,\,.
\label{DDXX1}
\ee
We have defined 
\begin{eqnarray}
\alpha_\pm =&& \frac {\pi\gamma_\pm}{\sqrt{\gamma_\pm +\rho^2}}\nonumber\\
\gamma_\pm^2 =&& (z_4{\rm sin\theta} - z_3 {\rm cos\theta} )^2 + (z_\perp-b\pm \frac d2)^2\nonumber\\
{\bf n}_+\cdot {\bf  n}_- =&&
\left((z_4\,{\rm sin}\theta - z_3\, {\rm cos}\theta )^2 + (b-z_\perp)^2 -\frac{d^2}4 \right)
\label{DDI5}
\end{eqnarray}
with ${\bf n}_\pm\cdot{\bf n}_\pm=\gamma_\pm^2$.
The scattering amplitude of an initial dipole through an instanton after
averaging over the global color orientation ${\bf R}$ is
\be
\frac 2{N_c}\,
\left( {\rm cos}\,\alpha_-\, {\rm cos}\, \alpha_+
+{\hat{\bf n}_-\cdot\hat{\bf n}_+}\,
{\rm sin}\,\alpha_-\,{\rm sin}\,\alpha_+
\right)\,{\bf 1}_{CD}
\label{ZZ1}
\ee
which reduces to the color-singlet channel. Specifically,
\be
{\cal W} (\theta , b) =\frac 2{N_c}\,\left(
{\rm cos}\,\alpha_-\, {\rm cos}\, \alpha_+\, 
+{\hat{\bf n}_-\cdot\hat{\bf n}_+}
\,\,{\rm sin}\,\alpha_-\,{\rm sin}\,\alpha_+\right)\,\,. \nonumber \\
\label{ZZ2}
\ee
The $\theta$ dependence 
in (\ref{ZZ1}-\ref{ZZ2}) can be readily eliminated by carrying the 
integration over the instanton position $z$ through the same change
of variable discussed in the quark-quark scattering, resulting in an amplitude
that depends only on $1/{\rm sin}\,\theta$. In Minkowski space this
translates to $1/s$ which vanishes at high energy. Indeed, the 
dipole-dipole scattering amplitude through a single instanton is
\be
\left<{\bf W} (\theta , b) \, {\bf W} (0,0) \right>\approx
\frac {n_0}{{\rm sin\, \theta}}\int\, d^4z\, \tilde{\cal W} (\theta , b) \, \tilde{\cal W} (0, 0) 
\label{DDI6}
\ee
where $\tilde{\cal W}$ follows from ${\cal W}$ by setting $\theta=\pi/2$.
Note that in this case ${\cal W} (0,0) =\tilde{\cal W} (0, 0)$.

It is clear from (\ref{DDXX1}) that while scattering through an
instanton, the dipole has to flip-color to keep track of the
velocity of the quarks in the dipole. The process is color-inelastic 
and therefore only contributes to the inelastic amplitude to first
order in the instanton density $n_0$, and to the elastic amplitude
to second order in the instanton density, a situation reminiscent
of one- and two-gluon exchange. 

The dipole-dipole scattering amplitude 
with open-color in the final state can be
constructed by using two dipole configurations
as given by (\ref{DDXX1}) with a relative angle $\theta$.
After averaging over the instanton color-orientations we
obtain
\be
&&{\cal W}_{AA}^{CD} (\theta, b)\,{\cal W}_{A'A'}^{C'D'} (0,0) =\nonumber\\&&
\frac 2{N_c}\,{\cal W}_1\,\,{\bf 1}_{CD}\,{\bf 1}_{C'D'} +\frac 1{N_c^2-1}\,
{\cal W}_{N_c^2-1}\,\,(\tau^a)_{DC}\,(\tau^a)_{D'C'}\,\,,\nonumber\\
\label{DDOPEN}
\ee
with the singlet part
\be
&&{\cal W}_1=
{\rm cos}\,\alpha_-\,{\rm cos}\,\alpha_+
{\rm cos}\,\underline\alpha_-\,{\rm cos}\,\underline\alpha_+\nonumber\\&& +
{\bf n}_-\cdot{\bf n}_+\,\,\underline{\bf n}_-\cdot\underline{\bf n}_+
{\rm sin}\,\alpha_-\,{\rm sin}\,\alpha_+\,{\rm sin}\,\underline{\alpha}_-\,{\rm sin}\,\underline{\alpha}_+\nonumber\\&&
+\underline{\bf n}_-\cdot\underline{\bf n}_+\,
{\rm cos}\,\alpha_-\,{\rm cos}\,\alpha_+
{\rm sin}\,\underline\alpha_-\,{\rm sin}\,\underline\alpha_+\nonumber\\&& 
+{\bf n}_-\cdot{\bf n}_+\,
{\rm sin}\,\alpha_-\,{\rm sin}\,\alpha_+
{\rm cos}\,\underline\alpha_-\,{\rm cos}\,\underline\alpha_+\,\,\,\,,
\label{DDSING}
\ee
and the octet part
\be
&&{\cal W}_{N_c^2-1}=
-{\rm cos}\,\alpha_-\,{\rm sin}\,\alpha_+
{\rm cos}\,\underline\alpha_-\,{\rm sin}\,\underline\alpha_+ {\bf n}_+\cdot\underline{\bf n}_+\nonumber\\&& 
-{\rm sin}\,\alpha_-\,{\rm cos}\,\alpha_+
{\rm sin}\,\underline\alpha_-\,{\rm cos}\,\underline\alpha_+ {\bf n}_-\cdot\underline{\bf n}_-\nonumber\\&& 
+{\rm cos}\,\alpha_-\,{\rm sin}\,\alpha_+
{\rm sin}\,\underline\alpha_-\,{\rm cos}\,\underline\alpha_+ {\bf n}_+\cdot\underline{\bf n}_-\nonumber\\&& 
+{\rm sin}\,\alpha_-\,{\rm cos}\,\alpha_+
{\rm cos}\,\underline\alpha_-\,{\rm sin}\,\underline\alpha_+ {\bf n}_-\cdot\underline{\bf n}_+\nonumber\\&& 
-{\rm sin}\,\alpha_-\,{\rm sin}\,\alpha_+
{\rm sin}\,\underline\alpha_-\,{\rm sin}\,\underline\alpha_+ 
\nonumber\\&&\times({\bf n}_-\cdot\underline{\bf n}_-\,\,
{\bf n}_+\cdot\underline{\bf n}_+ -
{\bf n}_-\cdot\underline{\bf n}_+
{\bf n}_+\cdot\underline{\bf n}_-)\,\,\,. 
\label{DDOCTET}
\ee
As in the case of quark-quark scattering, the (color)
elastic dipole-dipole amplitude scales as $1/{\rm sin}\,\theta$ 
and vanishes at high energy after analytical continuation.
However the (color) inelastic part of the amplitude is not.
After performing the shift of variables described before,
the $\theta$ dependence drops from all the angles $\alpha$.
There is a remaining $\theta$ dependence in the four
combinations ${\bf n}\cdot \underline{\bf n}$. In general,
the $\theta$ dependence in the latter is linear
in ${\rm sin}\,\theta$  or ${\rm cos}\,\theta$, and one may
worry that the last term in (\ref{DDOCTET}) may involve
higher powers of the trigonometric functions, which would 
yield to an unphysical cross section growing as $s$ after
analytical continuation. We have checked that this is not
the case, since
\be
{\bf n}_-\cdot\underline{\bf n}_-\,\,
{\bf n}_+\cdot\underline{\bf n}_+ -
{\bf n}_-\cdot\underline{\bf n}_+
{\bf n}_+\cdot\underline{\bf n}_-\rightarrow
d^2\left( z_2^2-{\rm cos}\,\theta\,z_3 z_4'\right)\nonumber
\ee
where $ z_4'$ is the new $z_4$ after the change of variable.
Moreover, the ${\rm cos}\,\theta$ term drops in the
integral over z (odd under $z_3\rightarrow -z_3$), making
this contribution to (\ref{DDOCTET}) subleading at high-energy
after analytical continuation~\footnote{This cancellation is not
generic. Indeed, the square of this contribution would be 
leading.}. Finally, we note that all ${\rm sin}\theta$ 
contributions in (\ref{DDOCTET}) drop following similar
parity considerations. As a result, the pertinent octet
contribution to the scattering amplitude is proportional
to ${\rm cotan}\,\theta$ which is $1/i{\rm tan}\,y=1/iv$
after analytical continuation.

We have assessed numerically the function  
\be \label{def_F8}
F_{N_c^2-1} \left(\frac b{\rho}_0, \frac d{\rho}_0\right)
=\frac {n_0}{{\rm cos}\,\theta} \int\, d^4 z  \,{\cal W}_{N_c^2-1}
\ee
which is shown in Fig.\,\ref{fig_dd_ampl} for different dipole sizes. We find that 
the dipole approximation scaling $F_{N_c^2-1}\sim d^2$
works well, even for sizes as large as the instanton size $d=\rho_0$.

\begin{figure}[t]
\includegraphics[width=3.in,angle=-90]{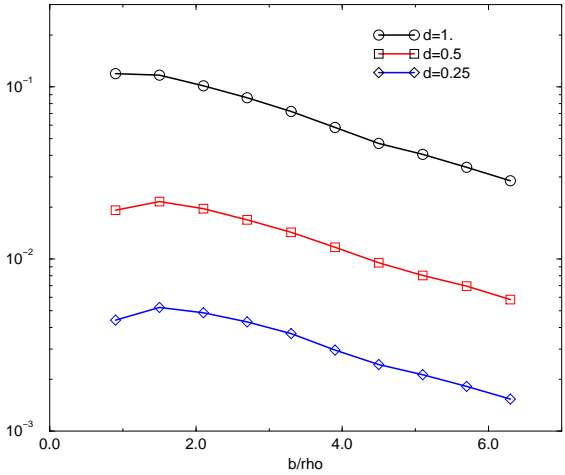}
\caption[]{
 \label{fig_dd_ampl}
(a) $F_{N_c^2-1}(b,d)$  defined in (\ref{def_F}) versus
the dipole-dipole
impact parameter b (in units of the instanton size $\rho_0$).
Each curve corresponds to a different dipole size $d$ (same units). 
}
\end{figure}

\section{Two-instanton effect}

We have shown above that the instanton contribution at large $s$
but small $t$ behaves in a way similar to one-gluon exchange: only
color-inelastic channels survive in the high energy limit. This means
that the contribution to the total cross section appears in the amplitude
squared, leading naturally to the concept of two-instanton exchange.
The latter contribution to each Wilson-line is more involved.
To streamline the discussion we will present the analysis of the two
instanton contribution to the differential cross-section of  quark-quark
scattering at high energy. Similar considerations apply to dipole-dipole
scattering as we briefly mention at the end of this section. Indeed,
for the quark-quark scattering, unitarity implies that the two-instanton 
 contribution to the differential cross section is
\begin{equation}
\frac{d\sigma}{dt} \approx \frac 1{s^2} \,\sum_{CD} \,\left| \,{\cal T}_{AC}^{BD}\,\right|^2\,\,,
\label{QQI5}
\end{equation}
with the averaging over the initial colors $A,B$ understood. 
Inserting (\ref{4}) after the substitution (\ref{QQI4}), we obtain
\begin{equation}
\frac{d\sigma}{dt} \approx \left(\frac {4n_0}{N_c}\right)^2 \,\int db\,db'\,e^{iq\cdot(b-b')}\,
\left({\bf J}+\frac 1{(N_c^2-1)}\, {\bf K}\right)
\label{QQI6}
\end{equation}
with 
\be
{\bf J}=&&\int\,d^4z\, ({\rm cos}\alpha -1)
({\rm cos}\underline{\alpha}-1)\,\nonumber\\&&\times\int\,d^4z'
({\rm cos}\alpha' -1)({\rm cos}\underline{\alpha}'-1)\nonumber\\
{\bf K}=&&\int\, d^4z\, \hat{\bf n}\cdot\underline{\hat{\bf n}}
\,\,{\rm sin}\,\alpha\,{\rm sin}\,\underline{\alpha}\nonumber\\&&\times
\int\, d^4z'\, \hat{\bf n}'\cdot\underline{\hat{\bf n}}'
\,\,{\rm sin}\,\alpha'\,{\rm sin}\,\underline{\alpha}'\,\,.
\label{QQI7}
\ee
The primed variables follow from the unprimed ones through
the substitution $z,b\rightarrow z',b'$. For large $\sqrt{s}$, 
${\bf J}\approx(1-F_{cc})(1-F_{cc}')/s^2$~\footnote{Up to self-energies.}
 and ${\bf K}=F_{ss}\,F_{ss}'$, so that
\be
\frac{d\sigma}{dt}\approx \frac{16\,n_0^2}{N_c^2(N_c^2-1)}\,
\left|\,\int\,db\,e^{iq\cdot b}\,F_{ss} \left(\frac b{\rho_0}\right)\right|^2\,\,.
\label{PARTIAL}
\ee
In particular,
the forward scattering amplitude in the two-instanton approximation is
\be
\sigma (t=0) \approx&&\frac {16\,n_0^2}{N_c^2 (N_c^2-1)}
\nonumber\\&&\times\int_0^\infty\,dq_\perp^2\,
\left|\,\int\,db\,e^{iq\cdot b}\,F_{ss} \left(\frac {b}{\rho_0}\right)\,\right|^2\,\,,
\ee
which is finite at large $\sqrt{s}$. Hence, for forward 
scattering partons in the instanton vacuum model, we have
\be
\sigma_{qq}\approx (n_0\,\rho_0^4)^2\,\rho_0^2
\ee

Clearly, the present analysis
generalizes to the dipole-dipole scattering amplitude by
using (\ref{DDOPEN}) instead of (\ref{QQI4}) and proceeding as
before. The outcome is a finite scattering cross section,
\be
&&\sigma (t=0) \approx\frac {4\,n_0^2}{(N_c^2-1)}
\nonumber\\&&\times\int_0^\infty\,dq_\perp^2\,
\left|\,\int\,db\,e^{iq\cdot b}\,F_{N_c^2-1} \left(\frac {b}{\rho_0},
\frac d{\rho_0}\right)\,\right|^2\,\,.
\ee
Generically, the dipole-dipole cross section relates to the
quark-quark cross-section in the forward direction through
\be
\sigma_{dd}\approx \sigma_{qq}\,\,\frac{(d_1d_2)^2}{\rho_0^4}\,\,.
\label{DDSECOND}
\ee

It is instructive to compare our instanton results
to those developed by Dosch and collaborators~\cite{DOSCH}
in the context of the stochastic vacuum model (SVM).
In brief, in the SVM model the Wilson-lines are expanded
in powers of the field-strength using a non-Abelian form
of Stokes theorem in the Gaussian approximation. A typical
hadronic cross section in the SVM model is
\be \label{dosch_sigma}
\sigma \approx \,\,<(gG)^2>^2 \,\,a^{10}\,\, {\bf F}\,(R_h/a) 
\label{DDSVM}
\ee
where the first factor is the ``gluon condensate'', $a$ is a
fitted correlation length, {\bf F} is some dimensionless
function depending on the hadronic radius $R_h$. 
%
Although our assumptions and those of \cite{DOSCH} are very
different regarding the character of the vacuum state, it is 
amusing to note the agreement between (\ref{DDSECOND})
and (\ref{DDSVM}). Indeed, the correlation length
$a$ of the SVM model is related (and in fact numerically
close) to our instanton radius
$\rho_0\approx 1/3$ fm, 
while the gluon condensate 
$<(gG)^2>$ 
 of the SVM model is simply proportional to the instanton density
 $n_0$
in the instanton model.
 
The most significant difference between these two approaches
apart from their dynamical content and the way we have carried
the analytical continuation, is the fact that we do not
expand in field strength. 	In fact, in the instanton model
there is no parameter which would allow to do so for strong
instanton fields. This difference is rather important as it is on
it that our conclusion regarding multiple color exchanges is based.  
(In the SVM model with Wick-theorem-like decomposition, those would 
be  just products of single exchanges, like in pQCD.)

\section{Cross section fluctuations} 

In so far, we have considered the $average$ value of the cross section
for a parton in a state of unit probability. However, 
partons and in general hadrons, are complex quantum mechanical states
~\footnote{A truly elementary particle 
 may have only one state and non-fluctuating cross section:
 it may have diffraction but no inelastic diffraction.}. Hence, the
quantum system is characterized by some amplitude of probability
through its wave function, and its corresponding scattering
cross section is probabilistic with a {\em probability
distribution} $P(\sigma)$. This idea was originally suggested 
by  Good and Walker \cite{GW}, who emphasized that
inelastic diffraction is a way to quantify this distribution 
via the second moment $\Delta\sigma^2= <(\sigma^2 -<\sigma>)^2>$. 

The extraction of this and the next (cubic) moment for the pion and
the nucleon using available data has been carried out years later
~\cite{xsectfluct} allowing for a reconstruction of the distribution $P(\sigma)$.
A striking aspect of these results is  that the 
nucleon fluctuations are large and comparable to
the pion fluctuations. This outcome does not fit
with the constituent quark model where the pion is a 
2-body system, and the nucleon is a 3-body system, with
more degrees of freedom. One of us \cite{Shu_toward}
had already noticed that this can be a further indication for
strongly correlated scalar diquarks in a nucleon. An experimental
test for this idea is to measure cross section
 fluctuations for a decuplet baryon 
such as $\Omega^-$. In the latter there are no  diquarks,
and smaller fluctuations (typical of a 3-body state) are expected. 
Another aspect of these fluctuations worth mentioning here is that
they seem to be maximal for $\sqrt{s}\approx 100$ GeV,
decreasing at very large energies. It
supports well the idea that the ``most
fluctuating'' partons are at $x\sim 10^{-2}$, while at much smaller x
one basically approaches a non-fluctuating black disk.

Although in the present paper we have limited our discussion
to issues of methodology, it is worth pointing out that the 
present concept of fluctuations in cross section can be used
to discriminate between the instanton effects herein described
and other descriptions based either on perturbative multi-gluon
exchange or non-perturbative vacuum structures. 

Indeed, the standard
multi-photon exchange in QED leads to an (eikonalized) exponential 
scattering amplitude, with Poisson-like fluctuations. If  
the mean-number of quanta exchanged $<n> \gg 1$ 
(e.g. for heavy ions with large Z$\approx 1/\alpha$), the distribution 
becomes narrow and we approach a classical limit, with weakly fluctuating 
scattering. Modulo color factors, the same conclusion applies to 
multi-gluon exchange in QCD. In contrast, 
the instanton-induced effects have completely different
statistical properties. The field of the instanton
itself is classical, hence coherent. However, the 
distribution over the instanton size and position is
quantum (in contrast to the Coulomb field of the ion
just mentioned), thereby leading to cross section fluctuations.
The latter are further enhanced by the
$diluteness$ of the instanton ensemble: the quark may appear very
black, provided a tunneling event happens to be close
to it, and rather transparent otherwise. As noticed already in ~\cite{Shu_82},
quarks are ``twinkling'' objects, as the associated  gauge/quark
fields
are strongly fluctuating.

To quantify some of these statements we show in Fig.(\ref{distr}) how such
 distribution looks like. We plot 
  $|F_{ss}(b=1)|^2$, at fixed impact parameter b=$\rho_0$. The
  distribution corresponds to instantons being homogeneously 
distributed in the 4d sphere around the center of the collision
point, with a radius $R_s\approx 2.2 \rho_0$ such that
$\pi^2R_s^4/2=1/n_0$, or in a smaller sphere within  $R< \rho_0$.  
However the resulting amplitude is highly inhomogeneous, with a large
peak at small amplitude, and a long tail at large amplitude.
Comparing  the solid and dashed curves, one can see that the latter is
due to instantons sitting near the center of the system. 

\begin{figure}[t]
  \epsfxsize=3.in
  \vspace{-.1in}
  \centerline{\epsffile{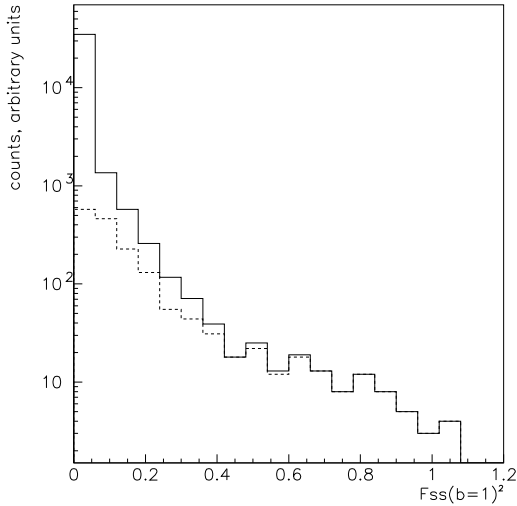}}
  \vspace{-.05in}
  \caption{\label{distr}
 The distribution $|F_{ss}(b=\rho_0)|^2$ with instantons filling
homogeneously
the  4d  `Wigner-Seitz sphere' of radius
2.2 $\rho_0$ (solid histogram) or the smaller sphere of radius
1.0 $\rho_0$ (dashed histogram).}
\end{figure}

\section{Conclusions and Outlook}
\subsection{Conclusions}

Several  new instanton-generated phenomena have been studied in
this work:  static potentials for color dipoles,  and high energy 
quark-quark and dipole-dipole scattering. The nature of the  
instanton effects makes their contribution to these processes 
different from the contribution expected in perturbation theory.

Overall,  the magnitude of the
instanton contribution is governed by two competing factors:
(i) a diluteness factor
$n_0\rho_0^4\ll 1$ reflecting the fact that
their density in the QCD vacuum  is small
($n_0\rho_0^4\ll 1$),
and (ii)  a classical enhancement factor, 
the instanton action which is large ($S_0/\hbar\approx 10\gg 1)$.
Naturally, the more partons are involved in a
particular process, the more powers of  $\alpha_s$ appear in the
perturbative result for a particular process. This penalty
does not apply to the instanton contribution.
One way to quantify this difference is to note that
the ratio of the instanton-to-perturbative contributions contains 
a power of the classical enhancement parameter, and this power grows 
with the number of partons involved. 
Typically the first power due to the classical instanton enhancement  
cannot really compensate for the small diluteness of the instantons in the 
vacuum. However, the second power is already 
sufficient to make  the instanton effects larger than the  perturbative ones
as we have now established for the potentials.
Indeed, the dipole-dipole instanton-induced potential exceeds 
significantly (by a factor $\sim$ 25)  the perturbative contribution
for distances $R>\rho_0$.

Based on these ideas, we have extended the analysis to near-forward
parton-parton scattering amplitudes, treating in details the case
of quark-quark and dipole-dipole scattering. Key to our analysis
was the concept of analytical continuation in the rapidity variable,
which we have applied to both the perturbative and instanton analysis
for comparison.

In the perturbative analysis, one- and two-gluon 
exchanges differ fundamentally in the sense that the former is 
color-changing (inelastic), while the latter is color-preserving
(elastic). Indeed, the two-gluon exchange 
mechanism~\cite{Low_Nussinov}  constitutes the starting ground for 
the soft pomeron approach to dipole-dipole scattering. Since the
instantons can be viewed as multi-gluon configurations (classical
fields), we have suggested that they maybe a viable starting point 
to analyze soft parton-parton scatterings. We have shown that the
instanton-induced amplitudes involve also color-elastic and 
color-inelastic channels. After analytical continuation, the 
one-instanton contribution to the color-elastic channel is purely
real and vanishes as $1/\sqrt{s}$ (much like a scalar exchange).
In other words,  in this work a single instanton is not ``cut'', its
multi-gluon content is not used.
 Instantons  contribute to soft parton-parton
scattering like the t-channel gluons mostly
through color exchange channels, or through re-scattering in the
elastic channel. The leading instanton contribution involves a 
two-instanton-prong channel, and yields a finite elastic
parton-parton scattering amplitude after analytical continuation
in rapidity space. Our result is reminiscent of the one reached
in the stochastic vacuum model~\cite{DOSCH}, although our assumptions
and methodology are different.

\subsection{Outlook}

The results we have derived were achieved in Euclidean space 
prior to our pertinent analytical continuation. Therefore, they
are testable from first principles by repeating our analysis 
using instead lattice QCD simulations. Indeed, the non-perturbative
dipole-dipole forces could be studied. In contrast to the 
quark-antiquark potential and to the best of our knowledge,
those forces have not been investigated on the lattice.
Also, the various scattering amplitudes discussed in the
present work  can and should be looked at, leading to
multi-parton amplitudes as we have qualitatively discussed.
Note, that not only the potentials
and scattering amplitudes themselves can be derived, but the degree of their
correlation with the presence of instantons in the underlying
configurations can be revealed as well, using lattice techniques
such as ``cooling'' and alike to help 
discriminate instantons by their topological charge.

Regarding the applications of our results, we 
admit that there remains a significant distance
to the description of real hadronic processes.
Although we hope to cover further phenomenological 
applications  elsewhere, we still would like to comment
on two broad but important dynamical issues: (i)
 the {\em mechanism of color rearrangements} in high energy collisions
and (ii) the issue of {\it hadronic substructure}
of the non-perturbative effects in the hadronic wave functions.

It is generally accepted that high energy hadronic processes can be
split into three  stages: (i) formation of hadronic wave function
(to which we turn later);
(ii) color re-arrangements of partons in a collision; and
(iii) decay of the arising system into multi-hadron final states.
It is further believed that at stage (iii) color flux tubes (QCD
strings)
are formed with basically $unit$ probability\footnote
{The fluxes are described   
by multiple phenomenological models/codes, e.g.  the Lund model.},
 so that one can ignore them in the calculation
of the cross section. Such assumption is implied in any perturbative
approach (such as the Low-Nussinov gluon-exchange model 
\cite{Low_Nussinov}), and we
assume the same is true for instanton-induced color exchanges as well.

\begin{figure}[ht]
\includegraphics[width=3.in]{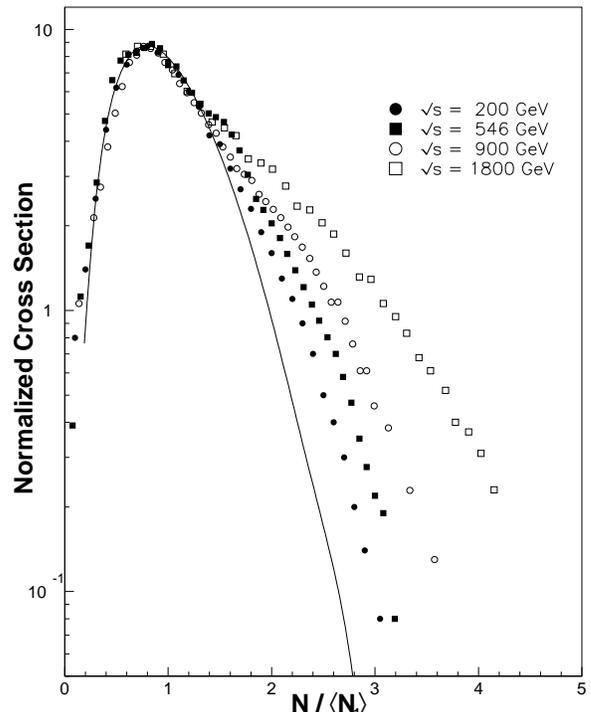}
  \caption{\label{fig_multiplicity}
Multiplicity distribution  in $\bar p p $ collisions, at 4 different 
energies from \protect\cite{MW}. At each energy the
cross section and multiplicity are rescaled, to put the low-N part at
the universal KNO curve (solid line). This is done to see better
the behavior of the ``second component'' discussed in the text.  }
\end{figure}

  Our main suggestion for further work is that
although the instanton-induced mechanism yields relatively small cross sections, this
mechanism is likely to dominate over  {\em events with  
multiple color rearrangements}. Is there experimental
evidence for this assertion in high-energy hadronic collisions?
An answer is provided by Fig.\,~\ref{fig_multiplicity} (taken from \cite{MW}),
which shows a (specially normalized) compilation of
multiplicity
distributions in $\bar p p$ collisions at various energies. 
The data shows that there is indeed  (at least) two components:
(i) one, with the cross section $\sigma_1(s)$
 and standard KNO distribution (well known from lower energy pp 
collisions), as indicated by
the solid curve; and (ii) another one with a different
  cross section $\sigma_2(s)$
and much higher multiplicity. Ascribing the main peak at
 $N/<N_1>\approx 0.8$ 
 to a $single$ color
rearrangement reaction (=2 QCD strings formed), one can conclude that at
the highest energy $\sqrt{s}=1800 GeV$ the multiplicity seen may
amount up to  10 strings.

The existence of the second component 
with double and triple multiplicity was anticipated by people
doing Regge theory decades ago, in the form of multi-pomeron exchanges. 
However, the multiplicity data shown in  Fig.\ref{fig_multiplicity} 
 do not really fit well into this description. The second components
simply does not look as  iterations of the first one.
There are no separate peaks and, more importantly, the s-dependence is
completely different. The first component is in fact consistent
with the approximation used above, namely
asymptotically constant  cross section 
(zero pomeron intercept), while the latter grows with $\sqrt{s}$ very strongly.

  Attempts to solve this puzzle in pQCD,
by summing ladder-type diagrams in leading log(x) approximation are well
known ~\cite{BFKL}, and they do indeed produce strongly growing 
cross sections and multi-parton states. So, the second component
may well be
due to those perturbative processes. 
 Non-perturbative
approaches (aiming mostly at the ``soft
pomeron'' or the first component discussed) have also been tried,
 from old fashion multi-peripheral
hadronic models (e.g. recent work \cite{Bj})
to mixed gluon-hadron ladders \cite{KL,Shu_toward}. Unfortunately, none
of these approaches have lead so far to a quantitative theory.

  Results/estimates made in this work lead to the conclusion,
  that instanton-induced color exchanges should dominate over  pQCD
t-channel gluons starting from the $double$ exchange amplitudes. It is
therefore logical to conjecture, that the second high-multiplicity component
   of pp collisions may be generated by this mechanism. That would
   explain
why multiple-string events are not just iteration of the first
component, and even consistent with where the transition
 appears to be.
Needless to say that much more work is still needed for a 
further test of this conjecture.

\vskip 1.5cm

{\bf Acknowledgements}
\vskip .35cm
We thank Maciek Nowak for discussions.
This work was supported in parts by the US-DOE grant 
DE-FG-88ER40388.

\end{narrowtext}
\end{document}